\documentclass{SciPost}

\hypersetup{
    colorlinks,
    linkcolor={red!50!black},
    citecolor={blue!50!black},
    urlcolor={blue!80!black}
}

\usepackage[bitstream-charter]{mathdesign}
\DeclareSymbolFont{usualmathcal}{OMS}{cmsy}{m}{n}
\DeclareSymbolFontAlphabet{\mathcal}{usualmathcal}

\fancypagestyle{SPstyle}{
\fancyhf{}
\lhead{\colorbox{scipostblue}{\bf \color{white} ~SciPost Physics }}
\rhead{{\bf \color{scipostdeepblue} ~Submission }}

\fancyfoot[C]{\textbf{\thepage}}
}

\usepackage{graphicx}
\usepackage[caption=false]{subfig}

\newcommand{\phantomsubfloat}[1]{
    {% apply caption setup only temporarily
        \captionsetup[subfigure]{labelformat=empty, labelfont=bf}
        \subfloat[][]{#1}
    }%
}

\usepackage{bm}

\begin{document}

\pagestyle{SPstyle}

\begin{center}{\Large \textbf{\color{scipostdeepblue}{
%%%%%%%%%% TODO: Write your article's title here
Free energy and metastable states in the square-lattice \\$\bm{J_1}$-$\bm{J_2}$ Ising model\\
}}}\end{center}

\begin{center}\textbf{
Veniamin A. Abalmasov
}\end{center}

\begin{center}
Institute of Automation and Electrometry SB RAS, 630090 Novosibirsk, Russia
\\[\baselineskip]
\href{mailto:abalmasov@iae.nsc.ru}{\small abalmasov@iae.nsc.ru}
\end{center}

\section*{\color{scipostdeepblue}{Abstract}}
\boldmath\textbf{Free energy as a function of polarization is calculated for the square-lattice $J_1$-$J_2$ Ising model for $J_2 < |J_1|/2$ using the random local field approximation (RLFA) and Monte Carlo (MC) simulations. Within RLFA, it reveals a metastable state with zero polarization in the ordered phase. In addition, the Landau free energy calculated within RLFA indicates a geometric slab-droplet phase transition at low temperature, which cannot be predicted by the mean field approximation. In turn, restricted free energy calculations for finite-size samples, exact and using MC simulations, reveal metastable states with a wide range of polarization values, but with only two domains. Taking into account the dependence of the restricted free energy on the nearest-neighbor correlations allows us to identify several more metastable states. The calculations also reveal additional slab-droplet transitions at $J_2 > |J_1|/4$. These findings enrich our knowledge of the $J_1$-$J_2$ Ising model and the RLFA as a useful theoretical tool to study phase transitions in spin systems. }

\vspace{\baselineskip}

%%%%%%%%%% BLOCK: Copyright information
% This block will be filled during the proof stage, and finilized just before publication.
% It exists here only as a placeholder, and should not be modified by authors.
\noindent\textcolor{white!90!black}{%
\fbox{\parbox{0.975\linewidth}{%
\textcolor{white!40!black}{\begin{tabular}{lr}%
  \begin{minipage}{0.6\textwidth}%
    {\small Copyright attribution to authors. \newline
    This work is a submission to SciPost Physics. \newline
    License information to appear upon publication. \newline
    Publication information to appear upon publication.}
  \end{minipage} & \begin{minipage}{0.4\textwidth}
    {\small Received Date \newline Accepted Date \newline Published Date}%
  \end{minipage}
\end{tabular}}
}}
}
%%%%%%%%%% BLOCK: Copyright information

% For convenience during refereeing we turn on line numbers:
%\linenumbers

\vspace{10pt}
\noindent\rule{\textwidth}{1pt}
\tableofcontents
\noindent\rule{\textwidth}{1pt}
\vspace{10pt}

% Write your article contents here, starting from first \section.
% An example structure is given below.

\section{Introduction}
\label{sec:intro}

The square-lattice $J_1$-$J_2$ Ising model is one of the minimal extension of the standard Ising model, in which the coupling $J_1$ between nearest neighbors is complemented by the coupling $J_2$ between diagonally next-nearest neighbors. The properties of this model are of both fundamental and practical interest, in particular, since its quantum Heisenberg counterpart is relevant to the antiferromagnetism in the parent compounds of the cuprate and pnictide families of high-temperature superconductors~\cite{si2016, dagotto1994, izyumov1997}. Indeed, recent state-of-the-art numerical calculations~\cite{lu2023, mai2022, jiang2021, jiang2020, jiang2019, huang2018,  huang2017, dodaro2017, jana2020, yu2014, lu2012} confirm earlier findings~\cite{husslein1996, szabo1997, hofstetter1998, huang2001, himeda2002, goswami2010, sentef2011, scalapino2012} that diagonal interactions are important in describing the available experimental data for these compounds. Spin frustrations due to the $J_2$ coupling affect the temperature and order of the phase transition and lead to a valence-bond-solid state in the transverse field Ising model~\cite{sadrzadeh2016, kellermann2019} and to a quantum spin liquid state in the Heisenberg model when $J_2$ is close to  $|J_1|/2$~\cite{jiang2012, li2012, wang2013, mustonen2018a}.

We recently highlighted the existence of metastable states with arbitrary polarization in the square-lattice $J_1$-$J_2$ Ising model in the interval $J_2 \in (0, |J_1|)$ using Monte Carlo (MC) simulations, which was further supported by simple microscopic energy considerations~\cite{abalmasov2023}. For the ferromagnetic ground state, i.e. for $J_1 < 0$ and $J_2 < |J_1|/ 2$, these states are rectangles with polarization opposite to the surrounding, briefly considered much earlier in~\cite{shore1991, shore1992}. Note that these states differ from the metastable states of the standard Ising model, consisting of straight stripes, into which a system with zero polarization, when applying the single-spin flip MC algorithm, relaxes after quenching only in about 1/3 of cases and only in the absence of an external field~\cite{spirin2001a, spirin2001b, olejarz2012}. Significantly, the random local field approximation~(RLFA)~\cite{vugmeister1987}, also applied in~\cite{abalmasov2023}, points to a metastable state with zero polarization in the same $J_2$ coupling range, thus reflecting the appearance of microscopic metastable states, which seems impossible for the mean field approximation~(MFA).

The polarization-dependent Landau free energy $F(m)$, considered in the framework of Landau's phenomenological theory of phase transitions~\cite{landau2013}, is an effective tool for studying metastable states (including those with polarization opposite to the external field) and can be used to calculate the relaxation rate of a system to the ground state using the Landau-Khalatnikov equation~\cite{landaulifshitz10} (see, e.g.,~\cite{abalmassov2013} for such calculations in ferroelectrics). It should be noted that for short-range interactions, the Landau free energy obtained within MFA differs qualitatively from the restricted free energy calculated for a fixed value of polarization exactly or using the MC method for finite-size samples~\cite{schulman1980, binder2007, binder2016}. In the former case,  below the phase transition temperature, the Landau free energy has a double-well shape. In the latter case, at a temperature close to zero,  the restricted free energy is determined by states with minimal energy, and the barrier between its two minima with opposite polarization corresponds to states with two large domains, the interface energy of which is proportional to the sample size~$L$. Thus, relative to the total energy, proportional to the number of spins $N = L^2$, the barrier vanishes in the thermodynamic limit~\cite{schulman1990, binder2011}. It was shown that, despite the loss of detailed information about microscopic spin configurations, the restricted free energy can be harnessed to well reproduce the MC polarization dynamics of the Ising model in good agreement with the droplet theory~\cite{lee1995}. These ideas were further developed in \cite{richards1996, richards1997, shteto1997, shteto1999}. The temperature dependence of the restricted free energy barrier in the three-dimensional $J_1$-$J_2$ Ising model was analytically estimated in~\cite{shore1992} in connection with domain growth and shrinking after low-temperature quenching.

Here, for the square-lattice $J_1$-$J_2$ Ising model, we calculate the Landau free energy in the RLFA framework and the restricted free energy exactly for a square sample of size $L = 6$ and using the MC method for $L = 10$ and $L = 14$. We pay special attention to the metastable states, which appear in this model and were studied earlier in~\cite{abalmasov2023}, and explore how they are reflected in the free energy. The features of the geometric slab-droplet phase transition in the free energy calculated by both methods are also briefly discussed.

\section{The $J_1$-$J_2$ Ising model}
\label{sec:model}

The square-lattice $J_1$-$J_2$ Ising model Hamiltonian reads
\begin{align}\label{hamiltonian}
   H = J_1 \sum_{\langle i, j \rangle} s_i s_j + J_2 \sum_{\langle \langle i, j \rangle \rangle} s_i s_j - \sum_i h_i s_i,
\end{align}
where each spin $s_i$ takes the value $+1$ or $-1$. The sums are over nearest $\langle i, j \rangle$ and diagonal next-nearest $\langle \langle i, j \rangle \rangle$ neighbors, as well as over each spin coupled to the external field $h_i$ at its position. In what follows, we set the values of the coupling constants $J_1 = -1$ and $J_2 < 1/2$, which correspond to the ferromagnetic ground state (the case $J_2 > 1/2$ with a striped antiferromagnetic ground state is similar in many aspects, but has a more complex spin topology and will be considered separately). Note that the model is invariant with respect to the simultaneous change of the sign of $J_1$ and the replacement of homogeneous polarization with N{\' e}el checkerboard one, corresponding to the antiferromagnetic order of the parent compounds of cuprate superconductors~\cite{lee2006}.

\section{Landau free energy in the random local field approximation}
\label{sec:RLFA}

RLFA is based on the exact formula for the average spin \cite{callen1963, vugmeister1987}:
\begin{align}\label{average-spin}
   \langle s_i \rangle = \langle \tanh \beta (h^s_i + h_i)\rangle,
\end{align}
where $\beta = 1/T$ is the inverse temperature in energy units. The local field, $h^s_i = - \sum_j J_{ij} s_j$, acting on the spin $s_i$ is caused by all spins $s_j$ coupled with it. 

The brackets in Eq.~(\ref{average-spin}) correspond to thermal averaging, which is performed with probability distribution~\cite{vugmeister1987, mertz2001}
\begin{align}   \label{probabilityM}
   P &= \prod _j (1 + m_j s_j)/2,
\end{align}
where the product is taken over all spins $s_j$ coupled to $s_i$, and $m_j = \langle s_j \rangle = m e^{i {\bf q} {\bf r}_j}$ is the thermally averaged polarization at position ${\bf r}_j$, the variation of which in space is determined by the propagation vector~${\bf q}$. The model applies to both ferromagnetic and ferroelectric materials, so we call $m$ polarization, which can be magnetic (magnetization) or electric. Here we consider only homogeneous polarization $m$ and external field $h$, which corresponds to ${\bf q} = (0, 0)$. Note that Eq.~(\ref{probabilityM}) implies that, within RLFA, the fluctuations of each spin are considered independent.

Eq.~(\ref{average-spin}) follows from equating to zero the derivative of the Landau free energy $F(m)$~\cite{lee1995}, which corresponds to thermodynamic equilibrium at a fixed value of polarization $m$~\cite{abalmassov2012b}. In order to reconstruct $F(m)$ and obtain its correct dependence on the external field $h$, we rewrite Eq.~(\ref{average-spin}) in the form $\partial \!F \!/\! \partial \!m \!=\! f(m) - h$, integration of which over $m$ yields $F(m) - F(0) \!=\! F_0(m)$, where $F_0(m) = \int_0^m \!f(m) dm -  h m$ is the Landau free energy minus the integration constant, which depends only on temperature.

\begin{figure*}[]
\centering
\includegraphics[width= 0.95 \textwidth]{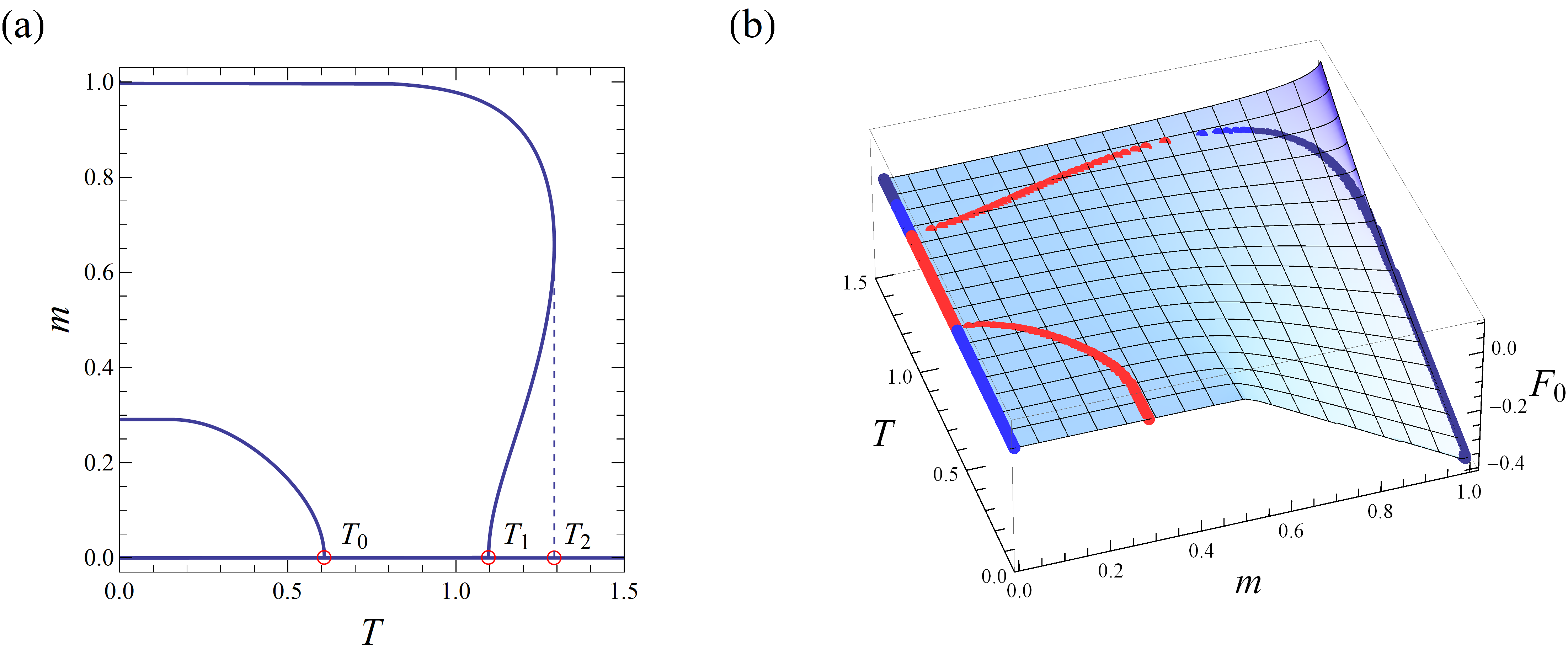}
\vspace{-1.0cm}
\phantomsubfloat{\label{fig:1a}}
\phantomsubfloat{\label{fig:1b}}
\caption{(a) The RLFA solution for $J_2 = 0.3$  (see also Fig.~2 in~\cite{abalmasov2023} for different values of $J_2$). (b) The Landau free energy $F_0(m)$ as a function of temperature within RLFA for $J_2 = 0.3$. Red dots correspond to the local maximum of $F_0(m)$ at each temperature (i.e. barrier), dark blue dots correspond to its global minimum (stable states), and light blue dots correspond to its local minimum (metastable states). A metastable state with zero polarization exists at temperatures from zero to $T_0$ and from $T_1$ to $T_c \approx 1.26 < T_2$, see also (a), at which a first order phase transition occurs according to RLFA.}
%\vspace{1.0 cm}
\label{fig:freeen_p03_3D}
\end{figure*}

\begin{figure}[]
\centering
\includegraphics[width= 0.45 \paperwidth]{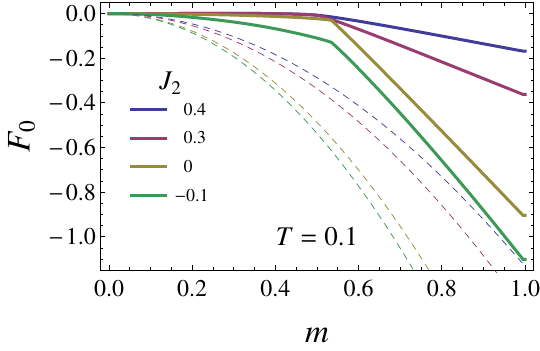}
\vspace{-0.cm}
\caption{The Landau free energy $F_0(m)$ calculated within RLFA (solid lines) for several values of $J_2$ at temperature $T = 0.1$, which shows a kink at polarization around $m = 0.5$. The dashed lines are the Landau free energies according to MFA.}
%\vspace{1.0 cm}
\label{fig:freeenergy_RLFA_T01}
\end{figure}

The RLFA solution and the Landau free energy $F_0(m)$ calculated in this way for $J_2 = 0.3$ are shown in Fig.~\ref{fig:freeen_p03_3D}. Here and in what follows, only values $m > 0$ are considered due to the symmetry $F(m) = F(-m)$ in the absence of an external field. The calculated free energy indeed points to the metastable state with $m = 0$, discussed in~\cite{abalmasov2023}. The dependence of the metastable state barrier on temperature and coupling constant $J_2$ is discussed in more detail in Appendix~\ref{app:A1}. It should be noted that within RLFA the transition turns out to be first order for $0. 25 \lesssim J_2 \lesssim 1.25$~\cite{abalmasov2023}, which is comparable to the cluster MFA results~\cite{jin2013, bobak2015, dominguez2021, krindges2023}, while recent more accurate calculations narrow this interval to a small region around $J_2 = 1/2$~\cite{jin2013, hu2021, li2021, yoshiyama2023, gangat2024}. RLFA also strongly underestimates the critical temperature $T_c$ just below $J_2 = 1/2$~\cite{abalmasov2023}, which could be attributed to increased frustration in this region. On the other hand, just above this value, where the ground state is striped antiferromagnetic, the accuracy of RLFA is quite high. It is also worth noting that the RLFA Landau free energy becomes flat in the limit~$J_2 \rightarrow 1/2$~(see Fig.~\ref{fig:freeenergy_RLFA_T01}), which indicates the absence of a phase transition at this point~\cite{abalmasov2023}.

At low temperature, the RLFA Landau free energy shows a kink at a polarization value $m_c \approx 0.5$ for $J_2 = 0.3$, see Fig.~\ref{fig:1b} and Fig.~\ref{fig:freeenergy_RLFA_T01}. A similar kink observed in the restricted free energy~\cite{berg1993, berg1993b, lee1995} is attributed to the geometrical phase transition~\cite{leung1990, moritz2017} and corresponds to the transition from a slab (for $m < m_c$), as the most probable configuration, to the droplet (for $m > m_c$), which will be discussed in the next section. For $J_2 = 0$, the RLFA prediction for $m_c \approx 0.53$  (Fig.~\ref{fig:freeenergy_RLFA_T01}) is close to the exact value of the critical polarization $m_c = 0.5$~\cite{leung1990}. We emphasize that RLFA is able to predict the geometrical phase transition, in contrast to MFA. We also checked that the mean field cluster approximation (even for cluster sizes up to $4 \times 4$ spins), formulated as in~\cite{blinc1966}, does not predict this transition, despite the good accuracy of the approximation in describing ferroelectric phase transitions~\cite{abalmassov2011, abalmassov2013b, abalmassov2016, abalmassov2019}. We can explain this as follows.  An essential part of obtaining the Landau free energy, as discussed above, is finding the dependence of the external field $h$ on the polarization $m$ using Eq.~(\ref{average-spin}). Within RLFA, the right-hand side of this equation is the sum of $\tanh$ functions with all possible values of the local field in their argument, while within MFA it contains only one term with the local field corresponding to the mean field. For different values of $m$, the solution of the equation corresponds to different local fields and, as a consequence, different spin configurations, as shown in Appendix~\ref{app:slab}.

\section{Restricted free energy calculated exactly for small samples}
\label{sec:L6}

Now we want to look at possible metastable states in the free energy of a sample of $N$ spins restricted to a certain value of the total spin $M$ and defined as~\cite{schulman1980, lee1995}
\begin{align}   \label{freeenergy_def}
   F (M) &= -T \log \sum_E n(M, E) \exp(-E/T),
\end{align}
where the sum is over all possible energies $E$ and $n(M, E)$ is the number of configurations for a given set of values, which would be the density of states in the continuous approximation. While the correspondence between polarization and total spin is obvious, $m = M / N$, here we prefer to consider the variable $M$, since it takes only integer values that are easy to relate with the corresponding spin configurations.

\begin{figure}[]
\centering
\includegraphics[width= 0.9 \textwidth]{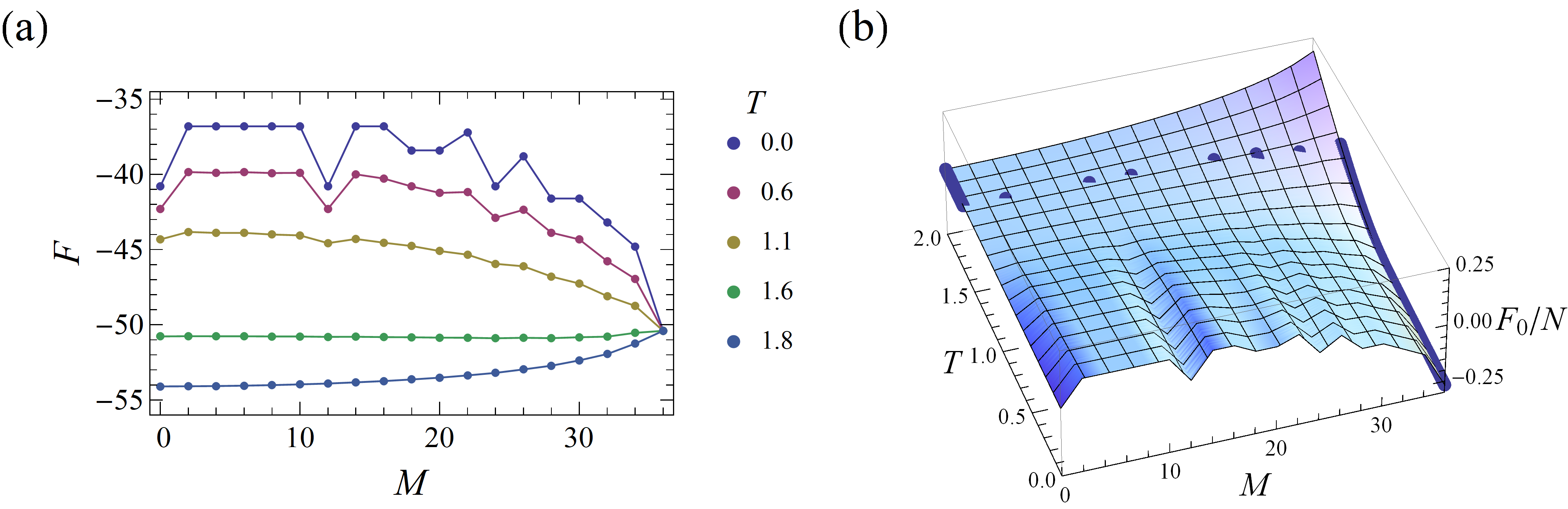}
\vspace{-0.cm}
\caption{(a) Restricted free energy $F$ calculated exactly by Eq.~(\ref{freeenergy_def}) at $J_2 = 0.3$ for sample size $L = 6$ as a function of total spin $M$ for a set of temperatures $T$ below and above the phase transition. (b) The same restricted free energy per spin $F_0/N$ as a function of total spin $M$ and temperature $T$. The free energy is defined only for integer even values of $M$ and linearly interpolated between them. For ease of comparison with the RLFA results in Fig.~\ref{fig:freeen_p03_3D}, we set here $F = 0$ at $M = 0$ at each temperature. Dark blue dots correspond to the global minimum of $F_0/N$ at each temperature.}
%\vspace{1.0 cm}
\label{fig:freeenergy_L6exact_p03_3D}
\end{figure}

For small samples, the sum in (\ref{freeenergy_def}) can be computed exactly. For a square sample of size  $L = 6$, yielding the total number of spins $N = 36$, the result for $J_2 = 0.3$ is shown in Fig.~\ref{fig:freeenergy_L6exact_p03_3D}. In all calculations we apply periodic boundary conditions (free boundary conditions are considered in Appendix~\ref{app:B2}). The critical temperature for $L = 6$ is equal to $T_c = 1.67$, while for $L = 100$ (which practically corresponds to an infinite sample size) it is $T_c = 1.26$~\cite{abalmasov2023}.

\begin{figure}[]
\centering
\includegraphics[width= 0.95 \columnwidth]{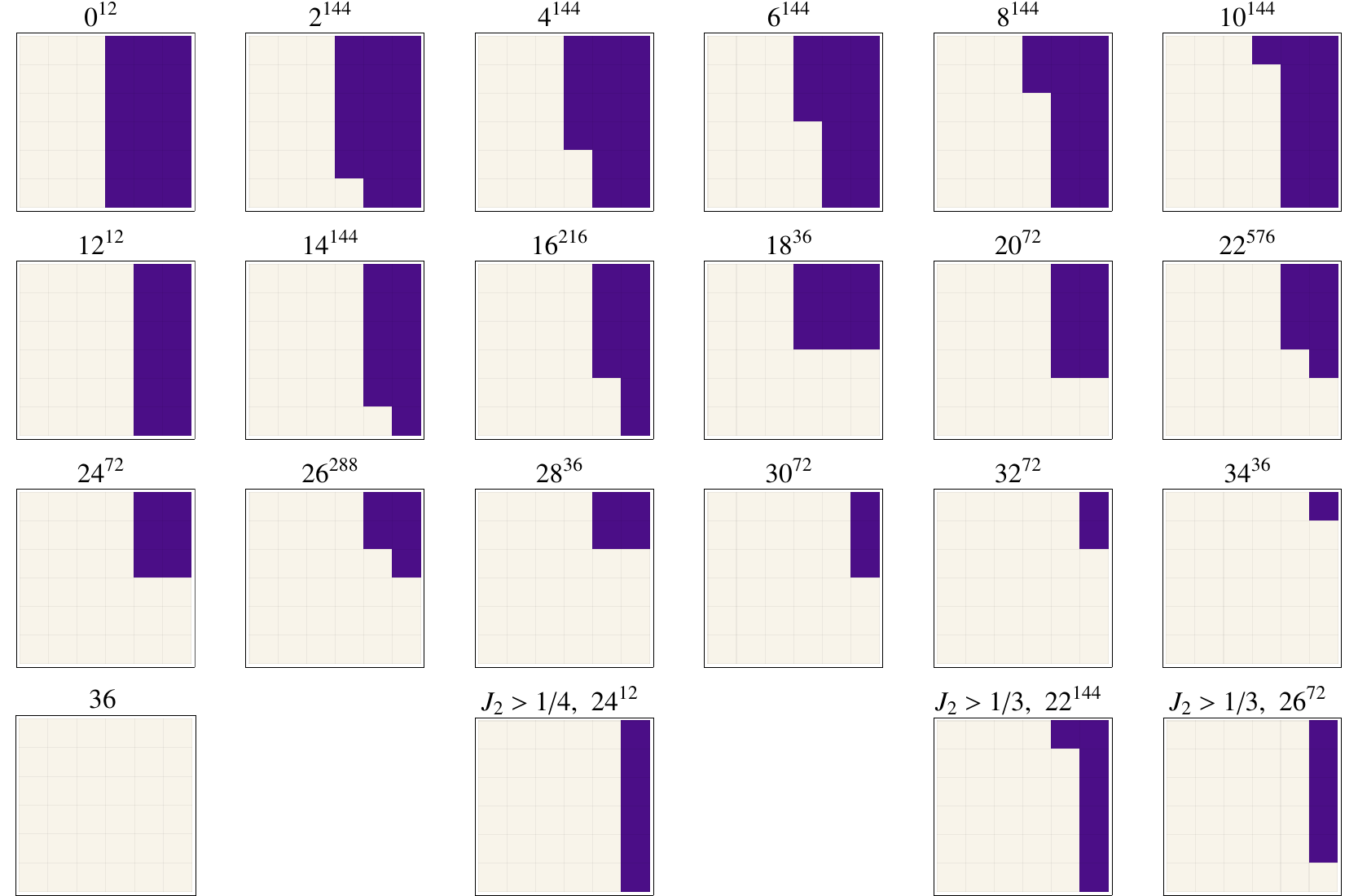}
\vspace{0.5cm}
\caption{Configurations, denoted $M^n$, that contribute to the restricted free energy $F(M)$ at zero temperature, i.e. have minimal energies for a given $M$, for all possible values of the total spin $M$ at $L = 6$; $n$ is their degeneracy due to translational and rotational symmetry. The configurations that change for $J_2 > 1/4$ and $J_2 > 1/3$ are shown separately in the bottom row (they affect the dependence of the energy barrier $\Delta F_M$ on $J_2$ as shown in Appendix~\ref{app:B}).}
%\vspace{1.0 cm}
\label{fig:config_L6}
\end{figure}

Configurations that contribute to the free energy $F(M)$ at zero temperature (i.e. have the lowest energies) for different values of $M$ are shown in Fig.~\ref{fig:config_L6}. At zero temperature, at $M < 16$, the restricted free energy is flat with a couple of pits, also mentioned in~\cite{lee1995}. These pits arise from spin configurations with a completely flat boundary between two slabs with opposite spin directions, see configurations with $M = 0$ ($M_{0}$) and $M = 12$ ($M_{12}$) in Fig.~\ref{fig:config_L6}. When a spin flips on this boundary (transitions $M_0 \rightarrow M_2$ and $M_{12} \rightarrow M_{14}$ in Fig.~\ref{fig:config_L6}) the energy increases by $4 J_1$. When the last spin in the row flips, the energy decreases by this value (transition $M_{10} \rightarrow M_{12}$ in Fig.~\ref{fig:config_L6}).  The distance between two neighbor pits of $F(M)$ is equal to $2 L$, since any single-spin flip changes the total moment by $\Delta M = \pm 2$. 

Metastable states inherent in the $J_1$-$J_2$ Ising model can be observed as local minima at $M = 18$, $20$, and $24$. It is interesting to note that while the transitions $M_{20} \rightarrow M_{22}$ and $M_{24} \rightarrow M_{26}$ are accompanied by an increase in energy, states $M_{18}$ and $M_{20}$ have the same energy, but are not connected to each other by a single spin flip. Thus, each of these two states are metastable in the sense of single-spin flip Monte Carlo dynamics. The same applies to state $M_{28}$ in Fig.~\ref{fig:config_L6}, which cannot be transformed to state $M_{30}$ with the same energy by a single-spin flip and is therefore metastable.

At $M \gtrsim N/2$, the configurations that contribute to the free energy $F(M)$ at zero temperature correspond mainly to the droplet (starting from configuration $M_{18}$ in Fig.~\ref{fig:config_L6}), but this depends on $J_2$, as indicated in the bottom row of Fig.~\ref{fig:config_L6}. At $J_2 > 1/4$ there is a reverse transition to the slab phase for $M_{24}$, and then back to the droplet phase for larger $M$. The same thing happens for $M_{22}$ when $J_2 > 1/3$. These threshold values of $J_2$ can be justified by explicitly calculating the energies of these states. We show the restricted free energy for various values of $J_2$ and the dependence of the metastable state barriers on $J_2$ and temperature in Appendix~\ref{app:B1}.

\begin{figure}[]
\centering
\includegraphics[width= 0.9 \textwidth]{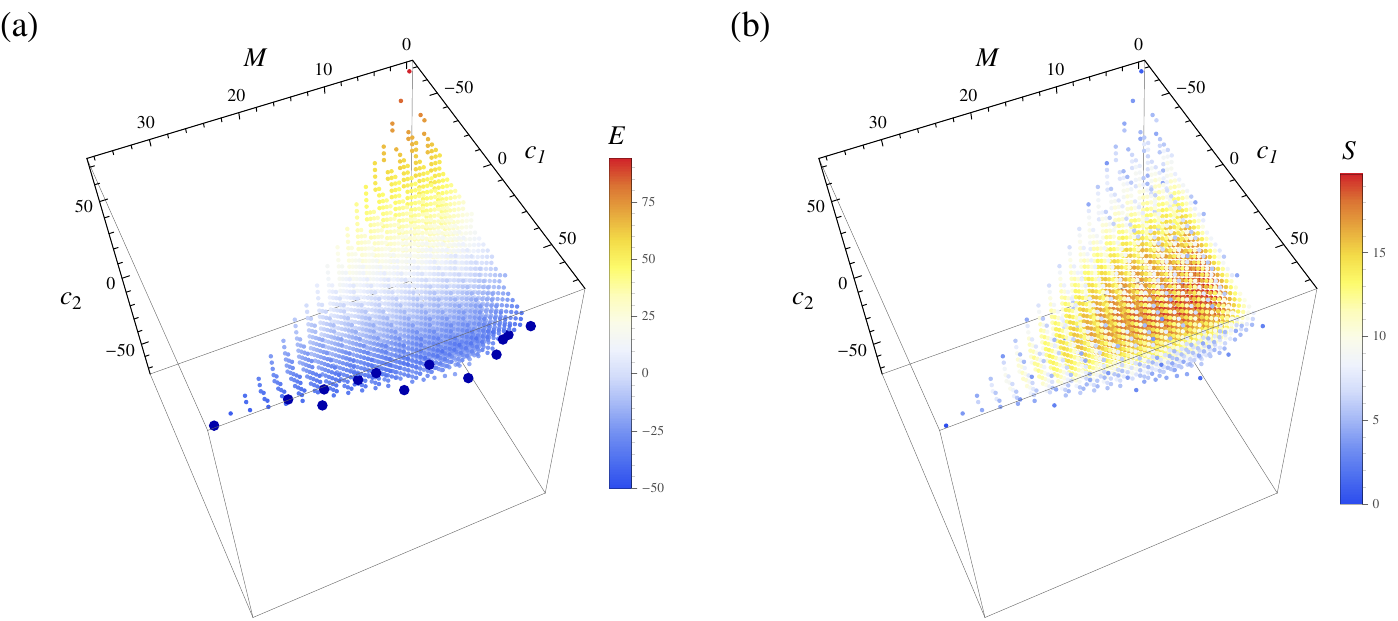}
\vspace{-0.cm}
\caption{(a) Energy $E$ as a color map for all possible states with total spin $M$ and nearest-neighbor correlations $c_1$ and $c_2$, (b) entropy, $S = \log n$, of these states. The restricted free energy is $F = E - T S$, where $T$ is the temperature. Large dark blue dots in (a) correspond to metastable states and one stable state with $M = 36$. }
%\vspace{1.0 cm}
\label{fig:ES_4D}
\end{figure}

Since the lowest energies that determine $F(M)$ at zero temperature imply a minimum number of domains in spin configurations (Fig.~\ref{fig:config_L6}), various metastable states with a large number of small domains of size at least $2 \times 2$, observed after low-temperature quenching in~\cite{abalmasov2023}, do not cause local minima of $F(M)$ (see Fig.~\ref{fig:freeenergy_L6exact_p03_3D} and Fig.~\ref{fig:config_L6}). However, we can try to reveal these states as local minima of the restricted free energy depending on the nearest spin correlation, e.g. such a function $F(M, c_1)$ with $c_1 = \sum_{\langle i, j \rangle}s_i s_j$ was calculated for the standard Ising model in~\cite{shteto1997}. This correlation can be easily and unambiguously calculated and is convenient for determining metastable states in the sense of Monte Carlo dynamics, since after a single-spin flip it changes only by $\Delta c_1$ equal to $0$, $\pm 4$, $\pm 8$. Here we calculate the restricted free energy $F(M, c_1, c_2)$, also taking into account the correlation with the next-nearest neighbor, $c_2 = \sum_{\langle\langle i, j \rangle\rangle}s_i s_j$. The result for the energy $E(M, c_1, c_2)$ and entropy, $S(M, c_1, c_2) = \log n$, which together constitute the free energy $F = E - T S$, is shown in Fig.~\ref{fig:ES_4D} as a color map. Note that in this case the energy is simply a linear function, since Eq.~(\ref{hamiltonian}) for the Hamiltonian becomes $H = J_1 c_1 + J_2 c_2 - h M$. In the absence of an external field ($h = 0$), it does not depend on $M$, but  possible correlations $c_{1,2}$ do.

\begin{figure}[]
\centering
\includegraphics[width= 0.9 \textwidth]{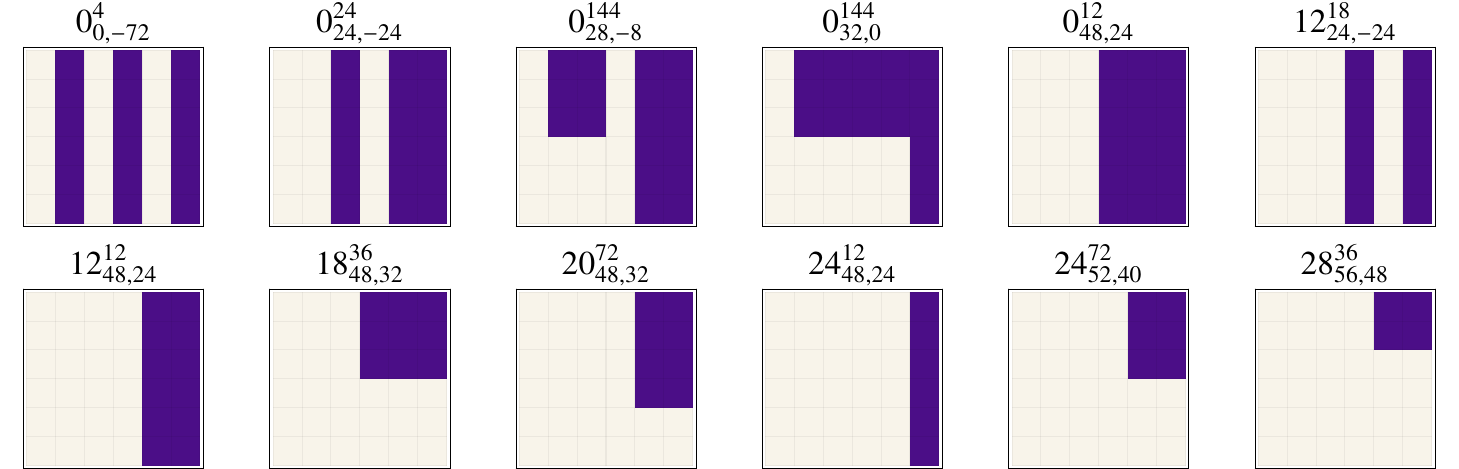}
\vspace{-0.cm}
\caption{Configurations of metastable states, denoted $M_{c_1, c_2}^n$, with total spin~$M$, nearest and next-nearest neighbor correlations~$c_1$ and~$c_2$, and number of configurations~$n$ for a given set of values.}
%\vspace{1.0 cm}
\label{fig:config_meta}
\end{figure}

Metastable states, the large dark blue dots in Fig.~\ref{fig:ES_4D}, were determined under the condition that any single-spin flip (with the only possible coordinate changes being $\Delta M = \pm 2$ and $\Delta c_{1,2} = 0$, $\pm 4$, $\pm 8$) will result in equal or greater energy. Thus, these states have the minimum possible energies for a given $M$ and turn out to be the same for all $J_2 \in (0, 0.5)$. Their configurations are shown in~Fig.~\ref{fig:config_meta}, which also includes configuration $M_{28}$ as discussed above. In addition to the metastable states for $F(M)$, we reveal for $F(M, c_1, c_2)$ only four higher-energy metastable states with $M = 0$ and one with $M = 12$. The metastability criterion used was too stringent to identify higher-energy metastable states that might not have all possible transitions $\Delta c_{1,2}$ that were tested for energy changes $\Delta E \geq 0$. Indeed, only a subset of states with a given set of values $(M, c_1, c_2)$ could be metastable, having different connectivity via a single-spin flip with other states, which seems impossible to take into account.

\section{Restricted free energy calculated for larger samples by Monte Carlo method}

For larger square samples, with $L = 7$ and $L = 8$, the free energy~(\ref{freeenergy_def}) can only be calculated using powerful supercomputers, given the large number of $2^N$ configurations for $N$ spins. Alternatively, it can be calculated approximately with sufficiently high accuracy using the Monte Carlo method. We use the Wang-Landau algorithm \cite{wang2001PRL, wang2001PRE, landau2004}, which has proven to be very efficient for this purpose at low temperature. It consists in performing a random walk in polarization and energy space to extract an estimate for the number of configurations $n(M, E)$ from Eq.~(\ref{freeenergy_def}) that produces a flat histogram. 

\begin{figure}[]
\centering
\includegraphics[width= 0.45 \paperwidth]{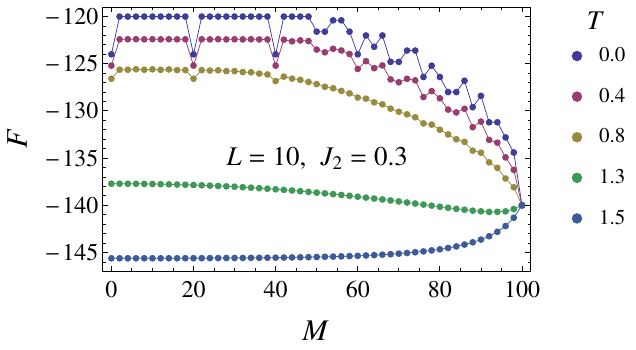}
\vspace{-0.cm}
\caption{Restricted free energy $F$ as a function of total spin $M$ for $J_2 = 0.3$ calculated by the MC method at several temperatures $T$ below and above the phase transition. The sample size is $L = 10$.}
%\vspace{1.0 cm}
\label{fig:freeenergy_L10_p03}
\end{figure}

Using the Wang-Landau algorithm, we reproduce the exact results for $L = 6$ with high accuracy and obtain similar results for $L = 10$, see Fig.~\ref{fig:freeenergy_L10_p03}, where a much larger number of metastable states are clearly visible at low temperature for $J_2 = 0.3$. Note that the restricted free energy for larger samples could also be calculated, as was done, for example, in \cite{lee1995} for $J_2 = 0$. However, for our purposes, namely to show how metastable states are reflected in the free energy, the size $L = 10$ seems optimal, and all metastable states are clearly visible and convenient for analysis. For larger samples, the total spin $M$ and free energy $F$ will scale as the number of spins $N$, but the height of the energy barriers will remain the same, i.e. $4 J_1$ (or half of it for free boundary conditions) for $M \lesssim N/2$ and $4 J_2$ for $M \gtrsim N/2$ (see Fig.~\ref{fig:freeenergy_L14_pAll} in App.~\ref{app:B2} for $L = 14$ and both periodic and free boundary conditions). The restricted free energy at $L = 10$ for various values of $J_2$ and the dependence of the metastable states barrier on $J_2$ and temperature are shown in Appendix~\ref{app:B1} and compared with the case $L = 6$.

The properties of the restricted free energy depending on nearest-neighbor correlations $F(M, c_1, c_2)$ is similar to the case $L = 6$. Using the same procedure described above for the case $L = 6$, several more metastable states can be identified. The energy of possible states and their entropy are shown in Appendix~\ref{app:B3}.

\section{Discussion}

The primary goal of this work was to consider the metastable states in the $J_1$-$J_2$ Ising model, recently found in the RLFA solution and MC simulations of low-temperature quenching~\cite{abalmasov2023}, from a free energy perspective. We found that the Landau free energy $F(m)$ as a function of polarization $m$ calculated within RLFA has a local minima at zero polarization below the ferromagnetic phase transition for $J_2 \in (0, 1/2)$ (Fig.~\ref{fig:freeen_p03_3D}), which indicates a metastable state. 

At the same time, the restricted free energy $F(M)$ as a function of the total spin $M$ calculated exactly for a small sample size $L = 6$ (Fig.~\ref{fig:freeenergy_L6exact_p03_3D}) and using MC simulations for $L = 10$ (Fig.~\ref{fig:freeenergy_L10_p03}) and larger (Fig.~\ref{fig:freeenergy_L14_pAll} in App.~\ref{app:B2}) has local minima corresponding to metastable states with various values of polarization. Some of them, namely at $M \lesssim N/2$, are due to long stripes with an activation energy of $4 J_1$ of a spin flip on a flat domain boundary~(Fig.~\ref{fig:config_L6}). In the standard Ising model, the system can become stuck in these states with a final polarization following a Gaussian distribution after zero-temperature quenching from an initially random configuration with zero polarization~\cite{spirin2001b}.

Metastable states at $M \gtrsim N/2$ are caused by droplet-shaped domains with an activation energy of $4 J_2$ of a spin flip in their corner, at least at $J_2 < 1/4$ for both sample sizes $L = 6$ and $L = 10$ (Fig.~\ref{fig:deltaF_L6_pAll} in App.~\ref{app:B1}).  At $J_2 > 1/4$, the dependence of the barrier height on $J_2$ changes (it is interesting whether this threshold value of $J_2$ holds true for any sample sizes). This corresponds to a change in the sequence of minimal energy configurations for increasing total spin $M$ (see Fig.~\ref{fig:config_L6}) with a return to the slab phase and then back again, which may be important for some applications. Indeed, the importance of the geometrical slab-droplet transition for various physical situations, including the dewetting transition between hydrophobic surfaces, was highlighted in~\cite{moritz2017}. 

As $J_2 \rightarrow 1/2$, some of the barriers disappear, while the rest tend to the value $4 |J_1|$ at a distance of $2L$ from each other~(Fig.~\ref{fig:freeenergy_L6_pAll} and Fig.~\ref{fig:deltaF_L6_pAll} in App.~\ref{app:B1}). The ground state with two slabs of different widths becomes extensively degenerate for large $L$~\cite{sadrzadeh2016, li2021} with $T_c = 0$~\cite{hu2021, lee2024}. However, when free boundary conditions are applied, the fully polarized state always represents the global energy minimum at zero temperature, even for $J_2 = 1/2$, when the energy gap is $2 |J_1|$ (see Fig.~\ref{fig:freeenergy_L14_pAll} in App.~\ref{app:B2}). This is due to the difference in the energy contributions of spins in the corners of domains compared to bulk spins. Note that the position of the slab-droplet phase transition is the same for both boundary conditions at $J_2 = 0$.

It should be noted that the metastable states into which the system relaxes after low-temperature quenching in~\cite{abalmasov2023} are not exactly the same as shown in Fig.~\ref{fig:config_L6} and which determines the free energy at zero temperature. Even considering nearest-neighbor correlations, we were only able to additionally reveal a few highly correlated metastable states in the restricted free energy (Fig.~\ref{fig:config_meta}). However, combining the metastable states shown in Fig.~\ref{fig:config_meta} with the proper domain spacing will also result in a metastable state, but with higher energy, bringing them closer to those observed in~\cite{abalmasov2023}. Since the energy of the metastable states in~\cite{abalmasov2023} is much higher, the system is more likely to get stuck in them, relaxing during quenching on the way to thermal equilibrium. Metastable states like in~Fig.~\ref{fig:config_meta} can in principle be reached after quenching at non-zero temperature after a sufficiently long relaxation time and domain coarsening, with a higher probability for those closer to the equilibrium polarization. At the same time, any of these states will be reached inevitably if the total spin is conserved during quenching, as in the Kawasaki~\cite{kawasaki1972} two-spin exchange algorithm, which is relevant for models describing transport phenomena caused by spatial inhomogenity such as diffusion, heat conduction, etc.

Although the zero polarization of the metastable state and the low height of the barrier proportional to the temperature near zero (see Fig.~\ref{fig:rlfa_freeen_barrier} in App.~\ref{app:A1}) is not exactly what follows from MC calculations, where the barrier heights are much higher and decrease with temperature (see Fig.~\ref{fig:deltaF_L6_T} in App.~\ref{app:B1}), the fact of even a rough indication of the metastable state by RLFA is very valuable. Another valuable RLFA prediction that turns out to be quite accurate is the geometric slab-droplet phase transition at zero temperature (Fig.~\ref{fig:1b} and Fig.~\ref{fig:freeenergy_RLFA_T01}). The reason why RLFA is so effective in this situation, in our opinion, is that by definition it takes into account the local field due to all possible configurations of spins interacting with the central spin, not just the mean field. The probability of these configurations, in turn, is determined by the polarization (see Appendix~\ref{app:slab}).

Finally, we will mention some recent advances in the experimental observation of meta-stable states using sub-picosecond optical pulses, which we believe could be applied to reveal metastable states discussed here and in~\cite{abalmasov2023}. For instance, in the quasi-two-dimensional antiferromagnet Sr$_2$IrO$_4$, a long-range magnetic correlation along one direction was converted into a glassy condition by a single 100-fs-laser pulse~\cite{wang2021prx}. Atomic-scale PbTiO$_3$/SrTiO$_3$ superlattices, counterpoising strain and polarization states in alternate layers, were converted by sub-picosecond optical pulses to a supercrystal phase in~\cite{stoica2019}. In a layered dichalcogenide crystal of $1T$-TaS$_2$, a hidden low-resistance electronic state with polaron reordering was reached as a result of a quench caused by a single 35-femtosecond laser pulse~\cite{stojchevska2014}. See also the references to relevant superconducting and magnetic materials with next-nearest-neighbor interactions mentioned in Introduction and~\cite{abalmasov2023}.

\section{Conclusion}

The Landau free energy calculated within RLFA for the square-lattice $J_1$-$J_2$ Ising model has a local minimum at zero polarization for $J_2 \in (0, |J_1|/2)$ at low temperature in the ferromagnetic state, indicating a metastable state. This reflects the appearance of local minima (metastable states) with various polarization values in the restricted free energy, which was calculated exactly and using MC simulations. The restricted free energy as a function of nearest-neighbor correlations shows several more metastable states, although all of them represent only the lowest energy part (with a minimum number of domains) of the metastable states observed after low-temperature quenching in~\cite{abalmasov2023}. We also show that RLFA predicts the slab-droplet phase transition as a kink in the polarization dependence of the Landau free energy at low temperature, and explain this by averaging over different spin configurations within RLFA. At the same time, the restricted free energy reveals additional slab-droplet transitions at $J_2 > |J_1|/4$, which may be important for applications.  Finally, we believe that metastable states should be taken into account when considering the low-temperature behavior of materials described by the $J_1$-$J_2$ Ising model. In addition, easy-to-use RLFA can help reveal the presence of metastable states and geometrical phase transitions in more complex systems, e.g., with site or bond disorder and spin tunneling in a transverse field.

\section*{Acknowledgments}

I thank B.E. Vugmeister for many useful discussions. The Siberian Branch of the Russian Academy of Sciences (SB RAS) Siberian Supercomputer Center is gratefully acknowledged for providing supercomputer facilities.

\paragraph{Funding information}
I acknowledge the support by the Ministry of Science and Higher Education of the Russian Federation (Grant No. 124041700106-2).

\begin{appendix}
\numberwithin{equation}{section}

\section{Landau free energy within RLFA}
\label{app:A}

\subsection{Metastable state barrier dependence on $J_2$ and $T$}
\label{app:A1}

\begin{figure*}[]
\centering
\includegraphics[width= 0.95 \textwidth]{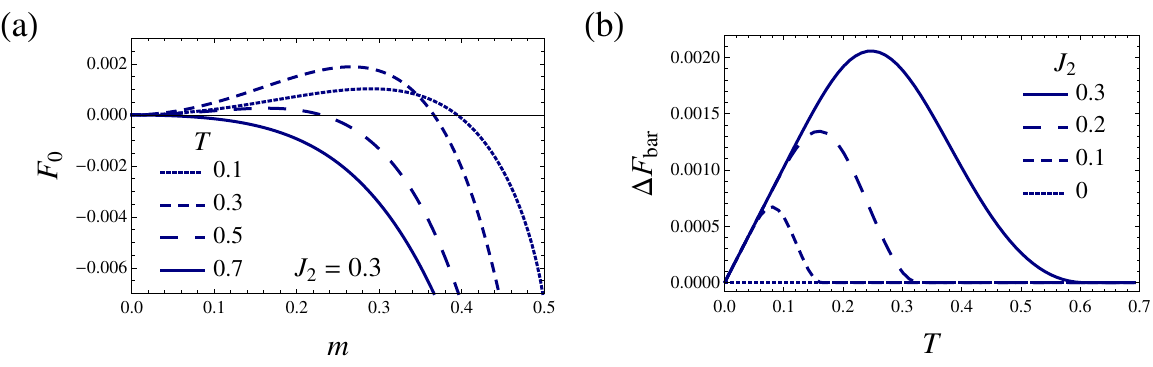}
\vspace{-1.cm}
\phantomsubfloat{\label{fig:2a}}
\phantomsubfloat{\label{fig:2b}}
\caption{(a) Restricted free energy $F_0(m)$ within RLFA for $J_2 = 0.3$ and polarization limited by $m \in (0,  0.5)$ to show the appearance at low temperature of a barrier at $m \neq 0$ whose height first increases and then decreases as the temperature approaches zero. (b) The barrier height $\Delta F_{\text{bar}} = F_{\text{bar}} - F(0)$ for the metastable state at $m = 0$, which appears in the restricted free energy $F(m)$ calculated within RLFA, as a function of temperature $T$. Only $J_2 < 0.31$ are considered, which implies $T_0 < T_1$ (for temperatures definition see Fig.~\ref{fig:1a}), since these two temperatures become undefined at larger $J_2$~\cite{abalmasov2023}.}
%\vspace{1.0 cm}
\label{fig:rlfa_freeen_barrier}
\end{figure*}

The metastable state with zero polarization (local minimum of the Landau free energy) is separated from the stable state by a barrier (local maximum), which appears within RLFA near $m = 0$ at temperature $T_0 \approx 0.6$ for $J_2 = 0.3$, see~Fig.~\ref{fig:freeen_p03_3D}. The barrier dependence on polarization $m$ and temperature for $J_2 = 0.3$ is shown in Fig.~\ref{fig:2a}. With decreasing temperature, the barrier height first increases and its position shifts up to $m \approx 0.29$, after which, at a temperature slightly less then $J_2$, the height begins to decrease linearly in $T$ to zero, see Fig.~\ref{fig:2b}. The maximum barrier height of about $0.002$ is close to the estimate in~\cite{abalmasov2023} based on the value of the coercive field. In Fig.~\ref{fig:2b}, the barrier height for various values of $J_2$ is shown. At $J_2 \approx 0.31$ we have $T_0 = T_1$, see Fig.~\ref{fig:freeen_p03_3D}, and for larger $J_2$ these two temperatures are not defined and the metastable RLFA solution $m = 0$ extends from zero to the critical temperature $T_c < T_2$.

\subsection{Details of the Landau free energy calculation}
\label{app:slab}

\begin{figure}[]
\centering
\includegraphics[width= 0.9 \textwidth]{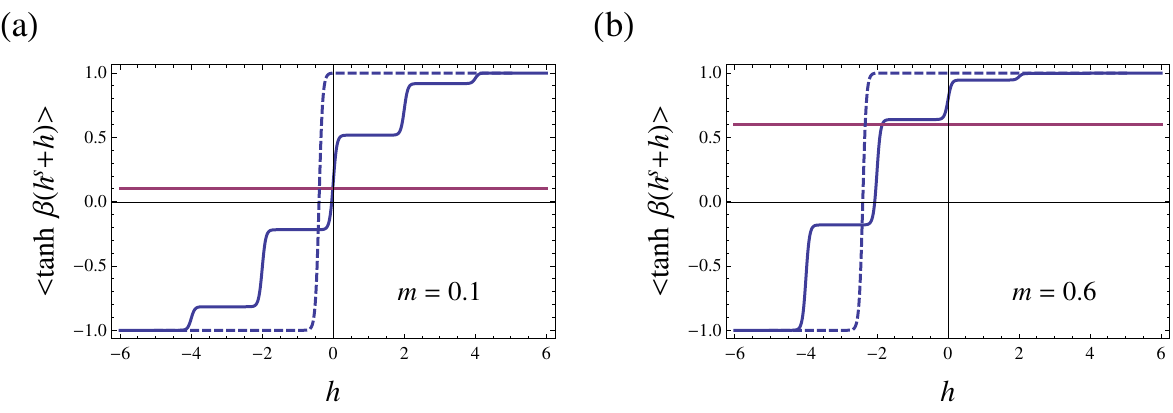} 
\vspace{-0.cm}
\caption{The right-hand side of the RLFA equation~(\ref{average-spin}) for $J_2 = 0$ and $T = 0.1$ (solid dark blue) as a function of the external field $h$ and its left-hand size (solid purple), which is just the polarization $m$. The intersection of two curves for a given value of $m$ yields the dependence $h(m)$. Dashed lines correspond to MFA.}
%\vspace{1.0 cm}
\label{fig:RLFA-equation}
\end{figure}

In order to obtain the Landau free energy from Eq.~\ref{average-spin}, we need to find from this equation the external field $h = f(m)$ as a function of the polarization $m$. This function is actually the derivative of the Landau free energy $\partial \!F \!/\! \partial \!m$, as discussed in Sec.~\ref{sec:RLFA}. Fig.~\ref{fig:RLFA-equation} shows the graphical solution of the Eq.~\ref{average-spin} with respect to $h$ for the Ising model (i.e. $J_2 = 0$) for $m = 0.1$ and $m = 0.6$ within both RLFA and MFA. The sum over different configurations of neighboring spins within RLFA yields the sum of $\tanh \beta(h^s + h)$ functions in Eq.~\ref{average-spin} with the corresponding local fields $h^s = - \sum_j J_{ij}s_j$ and weights equal to the probabilities of the configurations $P$ at a given value of $m$ (see Eq.~\ref{probabilityM}). Ultimately, this leads to a stepwise dependence $h(m)$ and, as a consequence, to a kink in $F(m)$ at a sufficiently low temperature. At the same time, within MFA the dependence $h(m)$ is smooth with no indication of the geometric phase transition.

\begin{figure}[]
\centering
\includegraphics[width= 0.9 \textwidth]{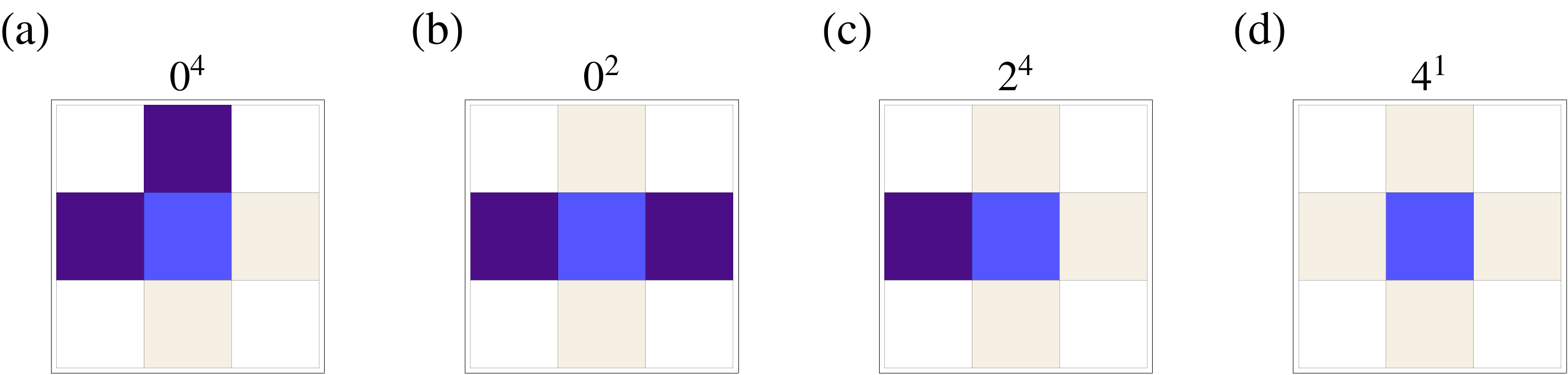}
\vspace{-0.cm}
\caption{In the Ising model ($J_1 = -1$, $J_2 = 0$), the local field $h^s = - \sum_{\langle j \rangle} J_{ij} s_j$ at the central spin position (in light blue) is determined by its four nearest neighbors with spin values of $+1$ (white) or $-1$ (dark blue). The value of the local field and the number of configurations $n$ due to rotational symmetry is indicated above each configuration as ${h^s}^n$. $h^s$ changes its sign when spins are flipped. Thus, the local field can have five different values when $J_2 = 0$.}
%\vspace{1.0 cm}
\label{fig:RLFA-config}
\end{figure}

\begin{figure}[]
\centering
\includegraphics[width= 0.9 \textwidth]{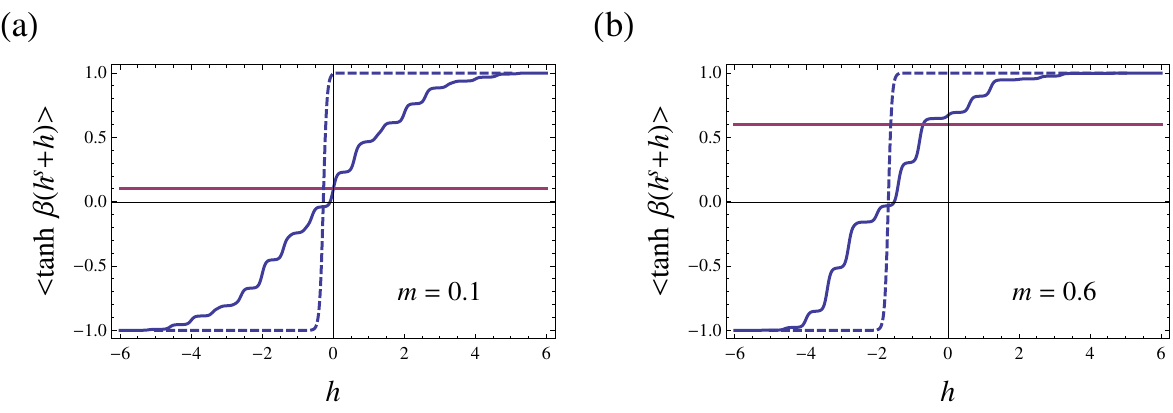} 
\vspace{-0.cm}
\caption{The right-hand side of the RLFA equation~(\ref{average-spin}) for $J_2 = 0.3$ and $T = 0.1$ (solid dark blue) as a function of the external field $h$ and its left-hand size (solid purple), which is just the polarization $m$. The intersection of two curves for a given value of $m$ yields the dependence $h(m)$. Dashed lines correspond to MFA.}
%\vspace{1.0 cm}
\label{fig:RLFA-equation-p03}
\end{figure}

For polarization $m$ from zero to about $1/2$, the external field $h$ is close to zero (Fig.~\ref{fig:RLFA-equation}(a)) and is determined by configurations that provide a zero local field $h^s$ (see Fig.~\ref{fig:RLFA-config}(a) and (b)). This corresponds to a central spin flip without a change in energy, and as a result, the Landau free energy is almost independent of $m$. This is similar to the behavior of the restricted free energy in the slab phase (see Sec.~\ref{sec:L6}), where a spin flip at the slab boundary does not change the energy.

At $m \gtrsim 1/2$, the most probable configurations of neighboring spins change (see Fig.~\ref{fig:RLFA-config}(c)) and yield the local field value of $h^s = 2$ (Fig.~\ref{fig:RLFA-equation}(b)). This implies that a further increase in $m$ will be accompanied by a minimization of domain boundaries and a decrease in $F(m)$, similar to the behavior of the restricted free energy in the droplet phase (Sec.~\ref{sec:L6}). This is how the slab-droplet phase transition is reflected within RLFA.

Note that the same mechanism is responsible for the predicting of the zero-polarization metastable state at $J_2 \in (0, 1/2)$. The number of local field values is squared in this case (see Fig.~\ref{fig:RLFA-equation-p03}), since for each local field contribution from the $J_1$ interaction (Fig.~\ref{fig:RLFA-config}) there are the same number of contributions from the $J_2$ coupling. Within RLFA, the $\tanh$ functions in Fig.~\ref{fig:RLFA-equation-p03} (and Fig.~\ref{fig:RLFA-equation}) do not move along the $h$ axis (see Eq.~\ref{average-spin}), but only their relative weights change with $m$ according to Eq.~\ref{probabilityM}. Thus, for small polarization $m$ in Fig.~\ref{fig:RLFA-equation-p03}(a), $h > 0$ (and hence the derivative of the Landau free energy $\partial \!F \!/\! \partial \!m > 0$), but at some value of $m$ the weight of configurations with negative local fields increases to such an extent that $h$ becomes zero again (which corresponds to the Landau free energy barrier) and then negative, as in Fig.~\ref{fig:RLFA-equation-p03}(b). Thus, the entire effect is due to the redistribution of weights of different local fields in Eq.~\ref{average-spin} (and the corresponding  configurations of the neighboring spins) as $m$ changes.

%\section{Restricted free energy calculated exactly for $L$ = 6}
\section{Restricted free energy properties}
\label{app:B}

\subsection{Metastable state barrier dependence on $J_2$ and $T$}
\label{app:B1}

\begin{figure*}[]
\centering
\includegraphics[width= 0.95 \textwidth]{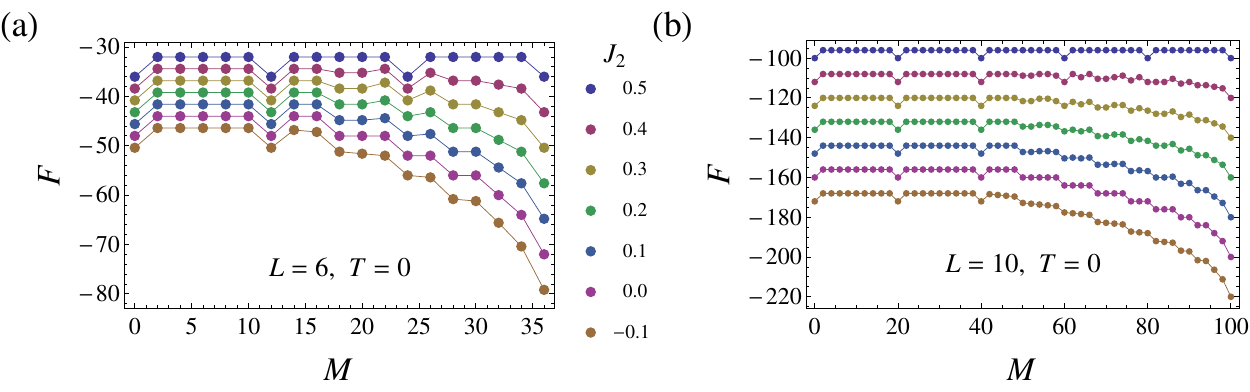}
\vspace{-1.cm}
\phantomsubfloat{\label{fig:6a}}
\phantomsubfloat{\label{fig:6b}}
\caption{Restricted free energy $F$ as a function of total spin $M$ at zero temperature $T = 0$, calculated according to Eq.~(\ref{freeenergy_def}) for several values of $J_2$ (listed in the legend and valid for both plots), (a) exactly for the sample size $L = 6$, (b) using the MC method for $L = 10$. The solid lines provide guides to the eye.}
%\vspace{1.0 cm}
\label{fig:freeenergy_L6_pAll}
\end{figure*}

\begin{figure*}[t]
\centering
\includegraphics[width= 0.95 \textwidth]{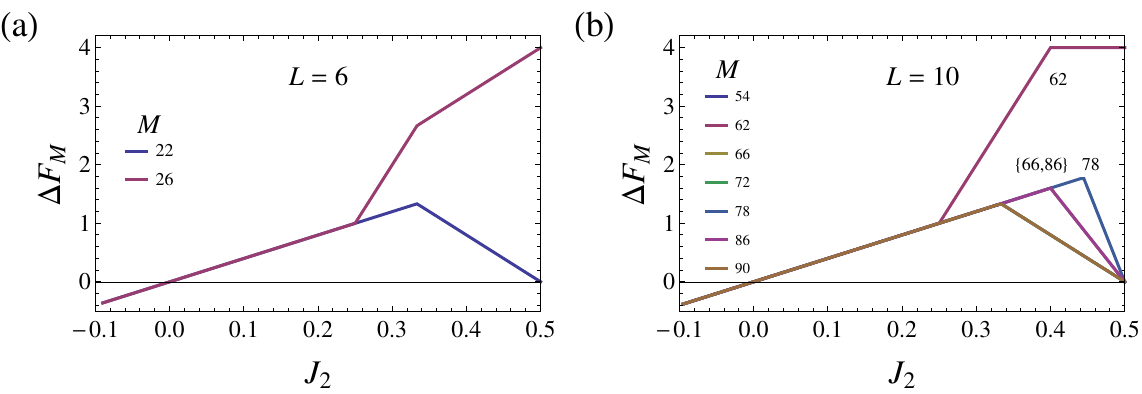}
\vspace{-1.cm}
\phantomsubfloat{\label{fig:7a}}
\phantomsubfloat{\label{fig:7b}}
\caption{Restricted free energy barrier height, $\Delta F_M = F_M - F_{M - 2}$, as a function of $J_2$ for several values of the total spin $M$ at zero temperature. (a) The sample size is $L = 6$. (b) The sample size is $L = 10$. Lines for different values of $M$ overlap. The lines corresponding to $M = 62, 78$ and overlaping $\{66, 86\}$ are marked separately.}
%\vspace{1.0 cm}
\label{fig:deltaF_L6_pAll}
\end{figure*}

In Fig.~\ref{fig:freeenergy_L6_pAll} we show side by side the restricted free energy at zero temperature for several values of $J_2$, from $-0.1$ up to $0.5$, calculated exactly for $L = 6$ and using the MC method for $L = 10$. It can be observed that the height of the barrier of metastable states, $\Delta F_M = F_M - F_{M - 2}$, depends on $J_2$. This is explicitly shown in Fig.~\ref{fig:deltaF_L6_pAll}. At $L = 6$, for $M = 22$ at $J_2 < 1/3$ and for $M = 26$ at $J_2 < 1/4$, the barrier height is $4 J_2$ and is determined by the spin flip at the corner of the droplet. The thresholds for $J_2$ follow from the energy ratio of the competing configurations for $M_{22}$, $M_{24}$, and $M_{26}$ in Fig.~\ref{fig:config_L6}. At $L = 10$, the barrier heights for metastable states in the droplet phase (i.e. at $M \gtrsim N/2$) also correspond to the spin flip at the corner of the droplet and are equal to $4 J_2$ for $J_2 <1/4$ (Fig.~\ref{fig:7b}). For $J_2 > 1/4$, this dependence changes due to slab-droplet and vice versa transitions, as in the case of $L = 6$. Some barriers, starting from a certain point, gradually disappear, others approach the barrier of the slab metastable state, equal to $4|J_1|$.

\begin{figure*}[t]
\centering
\includegraphics[width= 0.95 \textwidth]{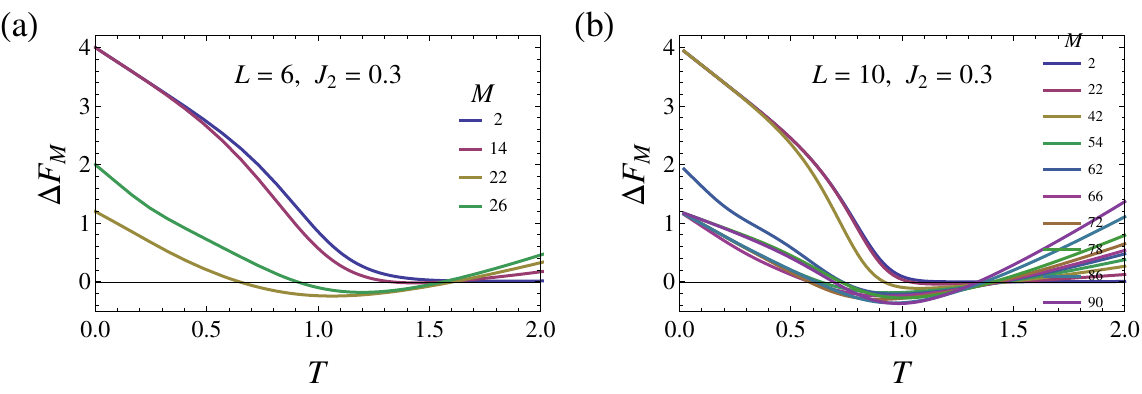}
\vspace{-1.cm}
\phantomsubfloat{\label{fig:8a}}
\phantomsubfloat{\label{fig:8b}}
\caption{Restricted free energy barrier height, $\Delta F_M = F_M - F_{M - 2}$, as a function of temperature for several values of the total spin $M$ and $J_2 = 0.3$. (a) The sample size is $L = 6$. (b) $L = 10$. The three upper curves correspond to $M = 2$, 22 and 42, and in the middle is $M = 62$.}
%\vspace{1.0 cm}
\label{fig:deltaF_L6_T}
\end{figure*}

At higher temperatures, other higher energy configurations in addition to those shown in Fig.~\ref{fig:config_L6} contribute to the partition function in Eq.~(\ref{freeenergy_def}) for each value of $M$. This affects the dependence of the above discussed energy barriers on temperature, which for $J_2 = 0.3$ is shown in Fig.~\ref{fig:deltaF_L6_T}. At $L = 6$, the metastable state barrier at $M = 22$ disappears at $T \approx 0.65$, which is close to the corresponding temperature $T_0 \approx 0.6$ from the RLFA solution (see Fig.~\ref{fig:freeen_p03_3D}). The case of $L = 10$ is very similar. We verified that the linear dependence of barriers height on temperature and their disappearance at a temperature close to $T_0$, obtained within RLFA, are also valid for other values of $J_2$. Note that the linear dependence on $T$ at low temperature follows directly from the definition of the free energy $F = E - T S$, where $E$ is the energy and $S$ is the entropy.

\subsection{Periodic vs free boundary conditions and larger samples}
\label{app:B2}

\begin{figure}[]
\centering
\includegraphics[width= 0.95 \textwidth]{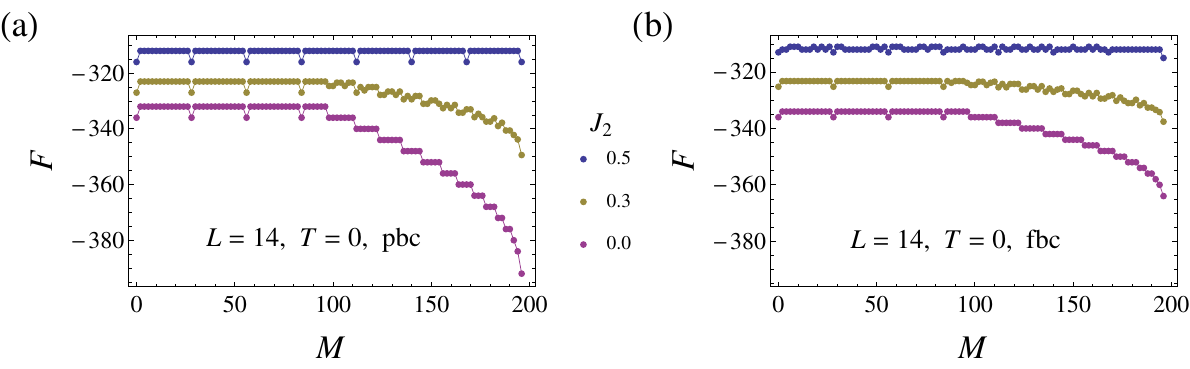}
\vspace{-0.cm}
\caption{Restricted free energy $F$ as a function of total spin $M$ at zero temperature $T = 0$, calculated according to Eq.~(\ref{freeenergy_def}) for several values of $J_2$ (listed in the legend and valid for both plots) using the MC method for $L = 14$, (a) for periodic boundary conditions, (b) for free boundary conditions. The curves for $J_2 = 0.3$ and $0.5$ are shifted down by  $75$ and $120$, respectively. The solid lines provide guides to the eye. }
%\vspace{1.0 cm}
\label{fig:freeenergy_L14_pAll}
\end{figure}

Changing the boundary conditions from periodic to free does not affect the restricted free energy at a qualitative level, but there are some differences, see Fig.~\ref{fig:freeenergy_L14_pAll}. In the latter case, the spin flip energy on a flat domain interface near the edge of the sample is $2 |J_1|$, which is half as much. This determines the energy barrier for metastable states in the slab phase and the decrease in energy when the entire side of the droplet dissolves in the droplet phase. At $J_2 = 0$, the sequence of configurations that determines the restricted free energy at zero temperature remains the same (Fig.~\ref{fig:config_L6}), but their degeneracy is less due to the lack of translational symmetry (two domains are always at the edge of the sample). The slab-droplet transition is present at the same value of $M$, but the dependence on $J_2$ is different. As a consequence, there are more local minima at $J_2 = 1/2$. The fully polarized state always represents the global energy minimum at zero temperature, even for $J_2 = 1/2$, when the energy gap is $2 |J_1|$. This is due to the difference in the energy contributions of spins in the corners of domains compared to bulk spins.

\subsection{Dependence on nearest-neighbor correlations for larger samples}
\label{app:B3}

\begin{figure}[]
\centering
\includegraphics[width= 0.9 \textwidth]{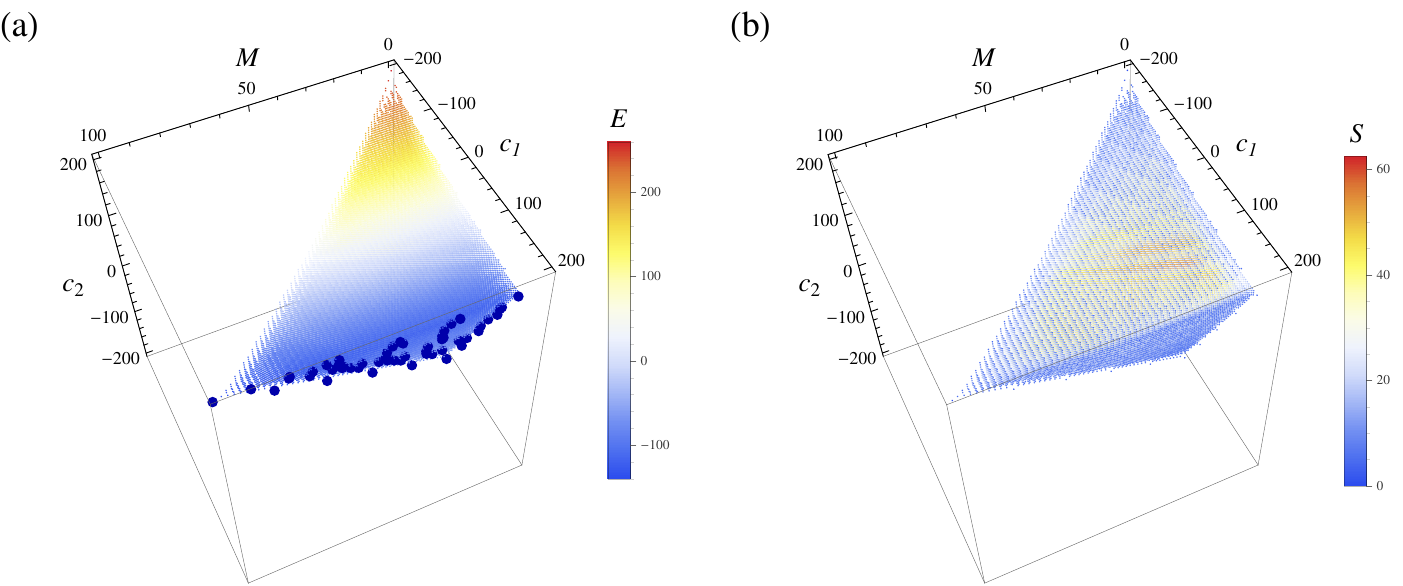}
\vspace{-0.cm}
\caption{Possible states with the total spin $M$ and nearest-neighbor correlations $c_1$ and $c_2$, and (a) their energy $E$ as a color map and (b) entropy $S = \log n$ for $L = 10$. The restricted free energy is $F = E - T S$, where $T$ is the temperature. Large dark blue dots correspond to metastable states and one stable state with $M = 100$. }
%\vspace{1.0 cm}
\label{fig:ES_L10_4D}
\end{figure}

As in the case of $L = 6$, we plot the energy $E(M, c_1, c_2)$ and entropy $S(M, c_1, c_2)$ as functions of the total spin $M$ and the nearest and next-nearest correlations ($c_1$ and $c_2$), which are calculated using MC simulations for the case of $L = 10$, see Fig.~\ref{fig:ES_L10_4D}. The restricted free energy is $F = E - T S$. In this case, the density of possible states is high and only states with extreme values of the variables close to the observer are clearly visible in the figure. Metastable states, marked by large dark blue dots, are found using the same procedure as in Sec.~\ref{sec:L6} for $L = 6$. They form a quasi-hypersurface in the states space, including only states with the lowest energies.

\end{appendix}

%\bibliography{../../biblioANNNI, ../../biblioAbalmasov}
%\bibliography{rlfa_J1J2_freeenergy_meta_v2_SP_resub2.bbl}

\begin{thebibliography}{10}
\providecommand{\url}[1]{\texttt{#1}}
\providecommand{\urlprefix}{URL }
\expandafter\ifx\csname urlstyle\endcsname\relax
  \providecommand{\doi}[1]{doi:\discretionary{}{}{}#1}\else
  \providecommand{\doi}{doi:\discretionary{}{}{}\begingroup
  \urlstyle{rm}\Url}\fi
\providecommand{\eprint}[2][]{\url{#2}}

\bibitem{si2016}
Q.~Si, R.~Yu and E.~Abrahams,
\newblock \emph{High-temperature superconductivity in iron pnictides and
  chalcogenides},
\newblock Nature Reviews Materials \textbf{1}(4), 16017 (2016),
\newblock \doi{10.1038/natrevmats.2016.17}.

\bibitem{dagotto1994}
E.~Dagotto,
\newblock \emph{Correlated electrons in high-temperature superconductors},
\newblock Reviews of Modern Physics \textbf{66}(3), 763 (1994),
\newblock \doi{10.1103/revmodphys.66.763}.

\bibitem{izyumov1997}
Y.~A. Izyumov,
\newblock \emph{{Strongly correlated electrons: the $t$-$J$ model}},
\newblock Physics-Uspekhi \textbf{40}(5), 445 (1997),
\newblock \doi{10.1070/pu1997v040n05abeh000234}.

\bibitem{lu2023}
X.~Lu, D.-W. Qu, Y.~Qi, W.~Li and S.-S. Gong,
\newblock \emph{{Ground-state phase diagram of the extended two-leg $t$-$J$
  ladder}},
\newblock Physical Review B \textbf{107}(12), 125114 (2023),
\newblock \doi{10.1103/physrevb.107.125114}.

\bibitem{mai2022}
P.~Mai, S.~Karakuzu, G.~Balduzzi, S.~Johnston and T.~A. Maier,
\newblock \emph{{Intertwined spin, charge, and pair correlations in the
  two-dimensional Hubbard model in the thermodynamic limit}},
\newblock Proceedings of the National Academy of Sciences \textbf{119}(7),
  e2112806119 (2022),
\newblock \doi{10.1073/pnas.2112806119}.

\bibitem{jiang2021}
S.~Jiang, D.~J. Scalapino and S.~R. White,
\newblock \emph{{Ground-state phase diagram of the $t$-$t'$-$J$ model}},
\newblock Proceedings of the National Academy of Sciences \textbf{118}(44),
  e2109978118 (2021),
\newblock \doi{10.1073/pnas.2109978118}.

\bibitem{jiang2020}
Y.-F. Jiang, J.~Zaanen, T.~P. Devereaux and H.-C. Jiang,
\newblock \emph{{Ground state phase diagram of the doped Hubbard model on the
  four-leg cylinder}},
\newblock Physical Review Research \textbf{2}(3), 033073 (2020),
\newblock \doi{10.1103/physrevresearch.2.033073}.

\bibitem{jiang2019}
H.-C. Jiang and T.~P. Devereaux,
\newblock \emph{{Superconductivity in the doped Hubbard model and its interplay
  with next-nearest hopping $t'$}},
\newblock Science \textbf{365}(6460), 1424 (2019),
\newblock \doi{10.1126/science.aal5304}.

\bibitem{huang2018}
E.~W. Huang, C.~B. Mendl, H.-C. Jiang, B.~Moritz and T.~P. Devereaux,
\newblock \emph{{Stripe order from the perspective of the Hubbard model}},
\newblock npj Quantum Materials \textbf{3}(1), 22 (2018),
\newblock \doi{10.1038/s41535-018-0097-0}.

\bibitem{huang2017}
E.~W. Huang, C.~B. Mendl, S.~Liu, S.~Johnston, H.-C. Jiang, B.~Moritz and T.~P.
  Devereaux,
\newblock \emph{{Numerical evidence of fluctuating stripes in the normal state
  of high-$T_c$ cuprate superconductors}},
\newblock Science \textbf{358}(6367), 1161 (2017),
\newblock \doi{10.1126/science.aak9546}.

\bibitem{dodaro2017}
J.~F. Dodaro, H.-C. Jiang and S.~A. Kivelson,
\newblock \emph{{Intertwined order in a frustrated four-leg $t-J$ cylinder}},
\newblock Physical Review B \textbf{95}(15), 155116 (2017),
\newblock \doi{10.1103/physrevb.95.155116}.

\bibitem{jana2020}
G.~Jana and A.~Mukherjee,
\newblock \emph{{Frustration effects at finite temperature in the half filled
  Hubbard model}},
\newblock Journal of Physics: Condensed Matter \textbf{32}(36), 365602 (2020),
\newblock \doi{10.1088/1361-648x/ab9058}.

\bibitem{yu2014}
R.~Yu, J.-X. Zhu and Q.~Si,
\newblock \emph{{Orbital-selective superconductivity, gap anisotropy, and spin
  resonance excitations in a multiorbital $t$-$J_1$-$J_2$ model for iron
  pnictides}},
\newblock Physical Review B \textbf{89}(2), 024509 (2014),
\newblock \doi{10.1103/physrevb.89.024509}.

\bibitem{lu2012}
X.~Lu, C.~Fang, W.-F. Tsai, Y.~Jiang and J.~Hu,
\newblock \emph{{s-wave superconductivity with orbital-dependent sign change in
  checkerboard models of iron-based superconductors}},
\newblock Physical Review B \textbf{85}(5), 054505 (2012),
\newblock \doi{10.1103/physrevb.85.054505}.

\bibitem{husslein1996}
T.~Husslein, I.~Morgenstern, D.~M. Newns, P.~C. Pattnaik, J.~M. Singer and
  H.~G. Matuttis,
\newblock \emph{{Quantum Monte Carlo evidence for d-wave pairing in the
  two-dimensional Hubbard model at a van Hove singularity}},
\newblock Physical Review B \textbf{54}(22), 16179 (1996),
\newblock \doi{10.1103/physrevb.54.16179}.

\bibitem{szabo1997}
Z.~Szab{\'{o}} and Z.~Gul{\'{a}}csi,
\newblock \emph{Superconductivity in the extended hubbard model with more than
  nearest-neighbour contributions},
\newblock Philosophical Magazine B \textbf{76}(6), 911 (1997),
\newblock \doi{10.1080/01418639708243138}.

\bibitem{hofstetter1998}
W.~Hofstetter and D.~Vollhardt,
\newblock \emph{{Frustration of antiferromagnetism in the $t$-$t'$-Hubbard
  model at weak coupling}},
\newblock Annalen der Physik \textbf{510}(1), 48 (1998),
\newblock \doi{10.1002/andp.19985100105}.

\bibitem{huang2001}
Z.~B. Huang, H.~Q. Lin and J.~E. Gubernatis,
\newblock \emph{{Quantum Monte Carlo study of Spin, Charge, and Pairing
  correlations in the $t$-$t'$-$U$ Hubbard model}},
\newblock Physical Review B \textbf{64}(20), 205101 (2001),
\newblock \doi{10.1103/physrevb.64.205101}.

\bibitem{himeda2002}
A.~Himeda, T.~Kato and M.~Ogata,
\newblock \emph{{Stripe States with Spatially Oscillating $d$-Wave
  Superconductivity in the Two-Dimensional $t$-$t'$-$J$ model}},
\newblock Physical Review Letters \textbf{88}(11), 117001 (2002),
\newblock \doi{10.1103/physrevlett.88.117001}.

\bibitem{goswami2010}
P.~Goswami, P.~Nikolic and Q.~Si,
\newblock \emph{{Superconductivity in multi-orbital t-J$_1$-J$_2$ model and its
  implications for iron pnictides}},
\newblock {EPL} (Europhysics Letters) \textbf{91}(3), 37006 (2010),
\newblock \doi{10.1209/0295-5075/91/37006}.

\bibitem{sentef2011}
M.~Sentef, P.~Werner, E.~Gull and A.~P. Kampf,
\newblock \emph{{Superconducting Phase and Pairing Fluctuations in the
  Half-Filled Two-Dimensional Hubbard Model}},
\newblock Physical Review Letters \textbf{107}(12), 126401 (2011),
\newblock \doi{10.1103/physrevlett.107.126401}.

\bibitem{scalapino2012}
D.~J. Scalapino and S.~R. White,
\newblock \emph{{Stripe structures in the $t$-$t'$-$J$ model}},
\newblock Physica C: Superconductivity \textbf{481}, 146 (2012),
\newblock \doi{10.1016/j.physc.2012.04.004}.

\bibitem{sadrzadeh2016}
M.~Sadrzadeh, R.~Haghshenas, S.~S. Jahromi and A.~Langari,
\newblock \emph{{Emergence of string valence-bond-solid state in the frustrated
  $J_1-J_2$ transverse field Ising model on the square lattice}},
\newblock Physical Review B \textbf{94}(21), 214419 (2016),
\newblock \doi{10.1103/physrevb.94.214419}.

\bibitem{kellermann2019}
N.~Kellermann, M.~Schmidt and F.~M. Zimmer,
\newblock \emph{{Quantum Ising model on the frustrated square lattice}},
\newblock Physical Review E \textbf{99}(1), 012134 (2019),
\newblock \doi{10.1103/physreve.99.012134}.

\bibitem{jiang2012}
H.-C. Jiang, H.~Yao and L.~Balents,
\newblock \emph{{Spin liquid ground state of the spin-$\frac{1}{2}$ square
  $J_1$-$J_2$ Heisenberg model}},
\newblock Physical Review B \textbf{86}(2), 024424 (2012),
\newblock \doi{10.1103/physrevb.86.024424}.

\bibitem{li2012}
Y.~Li, G.~Yu, M.~K. Chan, V.~Bal{\'{e}}dent, Y.~Li, N.~Bari{\v{s}}i{\'{c}},
  X.~Zhao, K.~Hradil, R.~A. Mole, Y.~Sidis, P.~Steffens, P.~Bourges
  \emph{et~al.},
\newblock \emph{{Two Ising-like magnetic excitations in a single-layer cuprate
  superconductor}},
\newblock Nature Physics \textbf{8}(5), 404 (2012),
\newblock \doi{10.1038/nphys2271}.

\bibitem{wang2013}
L.~Wang, D.~Poilblanc, Z.-C. Gu, X.-G. Wen and F.~Verstraete,
\newblock \emph{{Constructing a Gapless Spin-Liquid State for the Spin-$1/2$
  $J_1$ - $J_2$ Heisenberg Model on a Square Lattice}},
\newblock Physical Review Letters \textbf{111}(3), 037202 (2013),
\newblock \doi{10.1103/physrevlett.111.037202}.

\bibitem{mustonen2018a}
O.~Mustonen, S.~Vasala, E.~Sadrollahi, K.~P. Schmidt, C.~Baines, H.~C. Walker,
  I.~Terasaki, F.~J. Litterst, E.~Baggio-Saitovitch and M.~Karppinen,
\newblock \emph{Spin-liquid-like state in a spin-1/2 square-lattice
  antiferromagnet perovskite induced by d$^{10}$ {\textendash} d$^0$ cation
  mixing},
\newblock Nature Communications \textbf{9}(1), 1085 (2018),
\newblock \doi{10.1038/s41467-018-03435-1}.

\bibitem{abalmasov2023}
V.~A. Abalmasov and B.~E. Vugmeister,
\newblock \emph{{Metastable states in the $J_1$ - $J_2$ Ising model}},
\newblock Physical Review E \textbf{107}, 034124 (2023),
\newblock \doi{10.1103/PhysRevE.107.034124}.

\bibitem{shore1991}
J.~D. Shore and J.~P. Sethna,
\newblock \emph{{Prediction of logarithmic growth in a quenched Ising model}},
\newblock Physical Review B \textbf{43}(4), 3782 (1991),
\newblock \doi{10.1103/physrevb.43.3782}.

\bibitem{shore1992}
J.~D. Shore, M.~Holzer and J.~P. Sethna,
\newblock \emph{{Logarithmically slow domain growth in nonrandomly frustrated
  systems: Ising models with competing interactions}},
\newblock Physical Review B \textbf{46}(18), 11376 (1992),
\newblock \doi{10.1103/physrevb.46.11376}.

\bibitem{spirin2001a}
V.~Spirin, P.~L. Krapivsky and S.~Redner,
\newblock \emph{{Fate of zero-temperature Ising ferromagnets}},
\newblock Physical Review E \textbf{63}(3), 036118 (2001),
\newblock \doi{10.1103/physreve.63.036118}.

\bibitem{spirin2001b}
V.~Spirin, P.~L. Krapivsky and S.~Redner,
\newblock \emph{{Freezing in Ising ferromagnets}},
\newblock Physical Review E \textbf{65}(1), 016119 (2001),
\newblock \doi{10.1103/physreve.65.016119}.

\bibitem{olejarz2012}
J.~Olejarz, P.~L. Krapivsky and S.~Redner,
\newblock \emph{{Fate of 2D Kinetic Ferromagnets and Critical Percolation
  Crossing Probabilities}},
\newblock Physical Review Letters \textbf{109}(19), 195702 (2012),
\newblock \doi{10.1103/physrevlett.109.195702}.

\bibitem{vugmeister1987}
B.~E. Vugmeister and V.~A. Stephanovich,
\newblock \emph{{New random field theory for the concentrational phase
  transitions with appearance of long-range order. Application to the impurity
  dipole systems}},
\newblock Solid State Communications \textbf{63}(4), 323 (1987),
\newblock \doi{10.1016/0038-1098(87)90918-5}.

\bibitem{landau2013}
L.~D. Landau and E.~M. Lifshitz,
\newblock \emph{{Statistical Physics}}, vol.~5 of \emph{Course of Theoretical
  Physics},
\newblock Elsevier Science, Amsterdam, Netherlands,
\newblock ISBN 9780080570464 (2013).

\bibitem{landaulifshitz10}
E.~M. Lifshitz and L.~P. Pitaevskii,
\newblock \emph{{Physical Kinetics}}, vol.~10 of \emph{Course of theoretical
  physics},
\newblock Elsevier Science, Amsterdam, Netherlands,
\newblock ISBN 9780750626354 (2012).

\bibitem{abalmassov2013}
V.~A. Abalmassov, A.~M. Pugachev and N.~V. Surovtsev,
\newblock \emph{Dynamics of the order parameter and the potential of the
  hydrogen bond in a ferroelectric {DKDP} crystal},
\newblock Journal of Experimental and Theoretical Physics \textbf{116}(2), 280
  (2013),
\newblock \doi{10.1134/s1063776113020076}.

\bibitem{schulman1980}
L.~S. Schulman,
\newblock \emph{{Magnetisation probabilities and metastability in the Ising
  model}},
\newblock Journal of Physics A: Mathematical and General \textbf{13}(1), 237
  (1980),
\newblock \doi{10.1088/0305-4470/13/1/025}.

\bibitem{binder2007}
K.~Binder,
\newblock \emph{Double-well thermodynamic potentials and spinodal curves: how
  real are they?},
\newblock Philosophical Magazine Letters \textbf{87}(11), 799 (2007),
\newblock \doi{10.1080/09500830701496560}.

\bibitem{binder2016}
K.~Binder and P.~Virnau,
\newblock \emph{{Overview: Understanding nucleation phenomena from simulations
  of lattice gas models}},
\newblock The Journal of Chemical Physics \textbf{145}(21), 211701 (2016),
\newblock \doi{10.1063/1.4959235}.

\bibitem{schulman1990}
L.~S. Schulman,
\newblock \emph{{SYSTEM}-{SIZE} {EFFECTS} {IN} {METASTABILITY}},
\newblock In \emph{Finite Size Scaling and Numerical Simulation of Statistical
  Systems}, pp. 489--518. {WORLD} {SCIENTIFIC},
\newblock \doi{10.1142/9789814503419_0011} (1990).

\bibitem{binder2011}
K.~Binder, B.~Block, S.~K. Das, P.~Virnau and D.~Winter,
\newblock \emph{{Monte Carlo Methods for Estimating Interfacial Free Energies
  and Line Tensions}},
\newblock Journal of Statistical Physics \textbf{144}(3), 690 (2011),
\newblock \doi{10.1007/s10955-011-0226-7}.

\bibitem{lee1995}
J.~Lee, M.~A. Novotny and P.~A. Rikvold,
\newblock \emph{Method to study relaxation of metastable phases: Macroscopic
  mean-field dynamics},
\newblock Physical Review E \textbf{52}(1), 356 (1995),
\newblock \doi{10.1103/physreve.52.356}.

\bibitem{richards1996}
H.~L. Richards, M.~A. Novotny and P.~A. Rikvold,
\newblock \emph{{Analytical and computational study of magnetization switching
  in kinetic Ising systems with demagnetizing fields}},
\newblock Physical Review B \textbf{54}(6), 4113 (1996),
\newblock \doi{10.1103/physrevb.54.4113}.

\bibitem{richards1997}
H.~L. Richards, M.~Kolesik, P.-A. Lindg{\aa}rd, P.~A. Rikvold and M.~A.
  Novotny,
\newblock \emph{{Effects of boundary conditions on magnetization switching in
  kinetic Ising models of nanoscale ferromagnets}},
\newblock Physical Review B \textbf{55}(17), 11521 (1997),
\newblock \doi{10.1103/physrevb.55.11521}.

\bibitem{shteto1997}
I.~Shteto, J.~Linares and F.~Varret,
\newblock \emph{{Monte Carlo entropic sampling for the study of metastable
  states and relaxation paths}},
\newblock Physical Review E \textbf{56}(5), 5128 (1997),
\newblock \doi{10.1103/physreve.56.5128}.

\bibitem{shteto1999}
I.~Shteto, K.~Boukheddaden and F.~Varret,
\newblock \emph{{Metastable states of an Ising-like thermally bistable
  system}},
\newblock Physical Review E \textbf{60}(5), 5139 (1999),
\newblock \doi{10.1103/physreve.60.5139}.

\bibitem{lee2006}
P.~A. Lee, N.~Nagaosa and X.-G. Wen,
\newblock \emph{{Doping a Mott insulator: Physics of high-temperature
  superconductivity}},
\newblock Reviews of Modern Physics \textbf{78}(1), 17 (2006),
\newblock \doi{10.1103/revmodphys.78.17}.

\bibitem{callen1963}
H.~B. Callen,
\newblock \emph{{A note on Green functions and the Ising model}},
\newblock Physics Letters \textbf{4}(3), 161 (1963),
\newblock \doi{10.1016/0031-9163(63)90344-5}.

\bibitem{mertz2001}
D.~Mertz, F.~Celestini, B.~E. Vugmeister, H.~Rabitz and J.~M. Debierre,
\newblock \emph{{Coexistence of ferrimagnetic long-range order and cluster
  superparamagnetism in Li$_{1-x}$Ni$_{1+x}$O$_2$}},
\newblock Physical Review B \textbf{64}(9), 094437 (2001),
\newblock \doi{10.1103/physrevb.64.094437}.

\bibitem{abalmassov2012b}
V.~A. Abalmassov and A.~S. Yurkov,
\newblock \emph{Landau potential of a {KDP} crystal in the cluster
  approximation of the pseudospin model},
\newblock Physics of the Solid State \textbf{54}(5), 984 (2012),
\newblock \doi{10.1134/s1063783412050022}.

\bibitem{jin2013}
S.~Jin, A.~Sen, W.~Guo and A.~W. Sandvik,
\newblock \emph{{Phase transitions in the frustrated Ising model on the square
  lattice}},
\newblock Physical Review B \textbf{87}(14), 144406 (2013),
\newblock \doi{10.1103/physrevb.87.144406}.

\bibitem{bobak2015}
A.~Bob{\'{a}}k, T.~Lu{\v{c}}ivjansk{\'{y}}, M.~Borovsk{\'{y}} and
  M.~{\v{Z}}ukovi{\v{c}},
\newblock \emph{{Phase transitions in a frustrated Ising antiferromagnet on a
  square lattice}},
\newblock Physical Review E \textbf{91}(3), 032145 (2015),
\newblock \doi{10.1103/physreve.91.032145}.

\bibitem{dominguez2021}
E.~Dom{\'{\i}}nguez, C.~E. Lopetegui and R.~Mulet,
\newblock \emph{{Quantum cluster variational method and phase diagram of the
  quantum ferromagnetic $J_1-J_2$ model}},
\newblock Physical Review B \textbf{104}(1), 014205 (2021),
\newblock \doi{10.1103/physrevb.104.014205}.

\bibitem{krindges2023}
A.~Krindges, C.~V. Morais, M.~Schmidt and F.~M. Zimmer,
\newblock \emph{{Frustrated fermionic $J_1 - J_2$ model with pairing
  interaction}},
\newblock Journal of Magnetism and Magnetic Materials \textbf{577}, 170746
  (2023),
\newblock \doi{10.1016/j.jmmm.2023.170746}.

\bibitem{hu2021}
Y.~Hu and P.~Charbonneau,
\newblock \emph{{Numerical transfer matrix study of frustrated
  next-nearest-neighbor Ising models on square lattices}},
\newblock Physical Review B \textbf{104}(14), 144429 (2021),
\newblock \doi{10.1103/physrevb.104.144429}.

\bibitem{li2021}
H.~Li and L.-P. Yang,
\newblock \emph{{Tensor network simulation for the frustrated $J_1-J_2$ Ising
  model on the square lattice}},
\newblock Physical Review E \textbf{104}(2), 024118 (2021),
\newblock \doi{10.1103/physreve.104.024118}.

\bibitem{yoshiyama2023}
K.~Yoshiyama and K.~Hukushima,
\newblock \emph{{Higher-order tensor renormalization group study of the
  $J_1$-$J_2$ Ising model on a square lattice}},
\newblock Physical Review E \textbf{108}(5), 054124 (2023),
\newblock \doi{10.1103/physreve.108.054124}.

\bibitem{gangat2024}
A.~A. Gangat,
\newblock \emph{{Weak first-order phase transitions in the frustrated square
  lattice $J_1$-$J_2$ classical Ising model}},
\newblock Physical Review B \textbf{109}(10), 104419 (2024),
\newblock \doi{10.1103/physrevb.109.104419}.

\bibitem{berg1993}
B.~A. Berg, U.~Hansmann and T.~Neuhaus,
\newblock \emph{{Simulation of an ensemble with varying magnetic field: A
  numerical determination of the order-order interface tension in the $D$=2
  Ising model}},
\newblock Physical Review B \textbf{47}(1), 497 (1993),
\newblock \doi{10.1103/physrevb.47.497}.

\bibitem{berg1993b}
B.~A. Berg, U.~Hansmann and T.~Neuhaus,
\newblock \emph{{Properties of interfaces in the two and three dimensional
  Ising model}},
\newblock Zeitschrift für Physik B Condensed Matter \textbf{90}(2), 229
  (1993),
\newblock \doi{10.1007/bf02198159}.

\bibitem{leung1990}
K.~Leung and R.~K.~P. Zia,
\newblock \emph{Geometrically induced transitions between equilibrium crystal
  shapes},
\newblock Journal of Physics A: Mathematical and General \textbf{23}(20), 4593
  (1990),
\newblock \doi{10.1088/0305-4470/23/20/021}.

\bibitem{moritz2017}
C.~Moritz, A.~Tröster and C.~Dellago,
\newblock \emph{{Interplay of fast and slow dynamics in rare transition
  pathways: The disk-to-slab transition in the 2d Ising model}},
\newblock The Journal of Chemical Physics \textbf{147}(15), 152714 (2017),
\newblock \doi{10.1063/1.4997479}.

\bibitem{blinc1966}
R.~Blinc and S.~Svetina,
\newblock \emph{{Cluster Approximations for Order-Disorder-Type Hydrogen-Bonded
  Ferroelectrics. I. Small Clusters}},
\newblock Physical Review \textbf{147}(2), 423 (1966),
\newblock \doi{10.1103/physrev.147.423}.

\bibitem{abalmassov2011}
V.~A. Abalmassov, A.~M. Pugachev and N.~V. Surovtsev,
\newblock \emph{{Dielectric susceptibility of a deuterated {KDP} crystal from
  experiment on Raman scattering and in the cluster approximation}},
\newblock Physics of the Solid State \textbf{53}(7), 1371 (2011),
\newblock \doi{10.1134/s106378341107002x}.

\bibitem{abalmassov2013b}
V.~A. Abalmassov,
\newblock \emph{Landau coefficients and the critical electric field in a {KDP}
  crystal},
\newblock Bulletin of the Russian Academy of Sciences: Physics \textbf{77}(8),
  1012 (2013),
\newblock \doi{10.3103/s1062873813080030}.

\bibitem{abalmassov2016}
V.~A. Abalmassov,
\newblock \emph{Pressure effect on the ferroelectric phase transition in {KDP}
  in the cluster approximation of the proton-tunneling model},
\newblock Ferroelectrics \textbf{501}(1), 57 (2016),
\newblock \doi{10.1080/00150193.2016.1198976}.

\bibitem{abalmassov2019}
V.~A. Abalmassov,
\newblock \emph{{Monte Carlo studies of the ferroelectric phase transition in
  {KDP}}},
\newblock Ferroelectrics \textbf{538}(1), 1 (2019),
\newblock \doi{10.1080/00150193.2019.1569978}.

\bibitem{wang2001PRL}
F.~Wang and D.~P. Landau,
\newblock \emph{{Efficient, Multiple-Range Random Walk Algorithm to Calculate
  the Density of States}},
\newblock Physical Review Letters \textbf{86}(10), 2050 (2001),
\newblock \doi{10.1103/physrevlett.86.2050}.

\bibitem{wang2001PRE}
F.~Wang and D.~P. Landau,
\newblock \emph{Determining the density of states for classical statistical
  models: A random walk algorithm to produce a flat histogram},
\newblock Physical Review E \textbf{64}(5), 056101 (2001),
\newblock \doi{10.1103/physreve.64.056101}.

\bibitem{landau2004}
D.~P. Landau, S.-H. Tsai and M.~Exler,
\newblock \emph{{A new approach to Monte Carlo simulations in statistical
  physics: Wang-Landau sampling}},
\newblock American Journal of Physics \textbf{72}(10), 1294 (2004),
\newblock \doi{10.1119/1.1707017}.

\bibitem{lee2024}
J.~H. Lee, S.-Y. Kim and J.~M. Kim,
\newblock \emph{{Frustrated Ising model with competing interactions on a square
  lattice}},
\newblock Physical Review B \textbf{109}(6), 064422 (2024),
\newblock \doi{10.1103/physrevb.109.064422}.

\bibitem{kawasaki1972}
K.~Kawasaki,
\newblock \emph{{Kinetics of Ising Models}},
\newblock In C.~Domb and M.~S. Green, eds., \emph{{Phase Transitions and
  Critical Phenomena}}, vol.~2, chap.~11, pp. 443--501. Academic Press, London
  (1972).

\bibitem{wang2021prx}
R.~Wang, J.~Sun, D.~Meyers, J.~Q. Lin, J.~Yang, G.~Li, H.~Ding, A.~D. DiChiara,
  Y.~Cao, J.~Liu, M.~P.~M. Dean, H.~Wen \emph{et~al.},
\newblock \emph{{Single-Laser-Pulse-Driven Thermal Limit of the
  Quasi-Two-Dimensional Magnetic Ordering in Sr$_2$IrO$_4$}},
\newblock Physical Review X \textbf{11}(4), 041023 (2021),
\newblock \doi{10.1103/physrevx.11.041023}.

\bibitem{stoica2019}
V.~A. Stoica, N.~Laanait, C.~Dai, Z.~Hong, Y.~Yuan, Z.~Zhang, S.~Lei, M.~R.
  McCarter, A.~Yadav, A.~R. Damodaran, S.~Das, G.~A. Stone \emph{et~al.},
\newblock \emph{Optical creation of a supercrystal with three-dimensional
  nanoscale periodicity},
\newblock Nature Materials \textbf{18}(4), 377 (2019),
\newblock \doi{10.1038/s41563-019-0311-x}.

\bibitem{stojchevska2014}
L.~Stojchevska, I.~Vaskivskyi, T.~Mertelj, P.~Kusar, D.~Svetin, S.~Brazovskii
  and D.~Mihailovic,
\newblock \emph{{Ultrafast Switching to a Stable Hidden Quantum State in an
  Electronic Crystal}},
\newblock Science \textbf{344}(6180), 177 (2014),
\newblock \doi{10.1126/science.1241591}.

\end{thebibliography}
\bibliographystyle{SciPost_bibstyle}

\end{document}